\documentclass[11pt,preprint,showpacs,preprintnumbers,amsmath,amssymb]{revtex4}

\usepackage{exscale}
\usepackage{relsize}
\usepackage{graphicx}
\usepackage{dcolumn}
\usepackage{bm}
\usepackage{multirow}
\usepackage{CJK}
\usepackage{diagbox}
\usepackage{subfigure}

\begin{document}

\begin{CJK*}{GBK}{}

\title{Study of $CP$ Violation in $B^-\rightarrow K^- \pi^+\pi^-$ and $B^-\rightarrow K^- \sigma(600)$ decays in the QCD factorization approach}
\author{Jing-Juan Qi \footnote{e-mail: jjqi@mail.bnu.edu.cn}}
\affiliation{\small{College of Nuclear Science and Technology, Beijing Normal University, Beijing 100875, China}}

\author{Zhen-Yang Wang \footnote{e-mail: wangzhenyang@mail.nbu.edu.cn}}
\affiliation{\small{Physics Department, Ningbo University, Zhejiang 315211, China}}

\author{Xin-Heng Guo \footnote{Corresponding author, e-mail: xhguo@bnu.edu.cn}}
\affiliation{\small{College of Nuclear Science and Technology, Beijing Normal University, Beijing 100875, China}}

\author{Zhen-Hua Zhang \footnote{Corresponding author, e-mail: zhangzh@usc.edu.cn}}
\affiliation{\small{School of Nuclear and Technology, University of South China, Hengyang, Hunan 421001, China}}

\author{Chao Wang \footnote{e-mail: chaowang@nwpu.edu.cn}}
\affiliation{\small{Center for Ecological and Environmental Sciences, Key Laboratory for Space Bioscience and Biotechnology, Northwestern Polytechnical University, Xi'an 710072, China}}

\date{\today\\}

\begin{abstract}
In this work, we study the localized $CP$ violation in $B^-\rightarrow K^-\pi^+\pi^-$ and $B^-\rightarrow K^- \sigma(600)$ decays by employing the quasi two-body QCD factorization approach. Both the resonance and the nonresonance contributions are studied for the $B^-\rightarrow K^-\pi^+\pi^-$ decay. The resonance contributions include those not only from $[\pi\pi]$ channels including $\sigma(600)$, $\rho^0(770)$ and $\omega(782)$ but also from $[K\pi]$ channels including $K^*(892)$, $K_0^*(1430)$, $K^*(1410)$, $K^*(1680)$ and $K_2^*(1430)$. By fitting the experimental data $\mathcal{A_{CP}}(K^-\pi^+\pi^-)=0.678\pm0.078\pm0.0323\pm0.007$ for $m_{K^-\pi^+}^2<15$ $\mathrm{GeV}^2$ and $0.08<m_{\pi^+\pi^-}^2<0.66$ $\mathrm{GeV}^2$, we get the end-point divergence parameters in our model, $\phi_S \in [4.75, 5.95]$ and $\rho_S\in[4.2, 8]$. Using these results for $\rho_S$ and $\phi_S$, we predict that the $CP$ asymmetry parameter $\mathcal{A_{CP}} \in [-0.094, -0.034]$ and the branching fraction $\mathcal{B} \in [1.82, 20.0]\times10^{-5}$ for the $B^-\rightarrow K^-\sigma(600)$ decay. In addition, we also analyse contributions to the localized $CP$ asymmetry $\mathcal{A_{CP}}(B^-\rightarrow K^-\pi^+\pi^-)$ from $[\pi\pi]$, $[K\pi]$ channel resonances and nonresonance individually, which are found to be $\mathcal{A_{CP}}(B^-\rightarrow K^-[\pi^+\pi^-] \rightarrow K^-\pi^+\pi^-)=0.585\pm0.045$, $\mathcal{A_{CP}}(B^-\rightarrow [K^-\pi^+] \pi \rightarrow K^-\pi^+\pi^-)=0.086\pm0.021$ and $\mathcal{A_{CP}}^{NR}(B^-\rightarrow K^-\pi^+\pi^-)=0.061\pm0.0042$, respectively. Comparing these results, we can see that the localized $CP$ asymmetry in the $B^-\rightarrow K^-\pi^+\pi^-$ decay is mainly induced by the $[\pi\pi]$ channel resonances while contributions from the $[K\pi]$ channel resonances and nonresonance are both very small.
\end{abstract}


\maketitle
\end{CJK*}

\section{Introduction}

Nonleptonic decays of hadrons containing a heavy quark play an important role in testing the Standard Model (SM) picture of the Charge-Parity ($CP$) violation mechanism in flavor physics, improving our understanding of nonperturbative and perturbative QCD and exploring new physics beyond the SM. $CP$ violation is related to the weak complex phase in the Cabibbo-Kobayashi-Maskawa (CKM) matrix, which describes the mixing of different generations of quarks \cite{Cabibbo:1963yz, Kobayashi:1973fv}. Besides the weak phase, a large strong phase is also needed for a large \emph{CP} asymmetry. Generally, this strong phase is provided by QCD loop corrections and some phenomenological models.

Three-body decays of heavy mesons are more complicated than the two-body case as they receive resonant and nonresonant contributions and involve three-body matrix elements. The direct nonresonant three-body decay of mesons generally receives two separate contributions: one from the point like weak transition and the other from the pole diagrams that involve three-point or four-point strong vertices. The nonresonant background in charmless three-body B decays due to the transition $B\rightarrow M_1M_2M_3$ has been studied extensively  based on Heavy Meson Chiral Perturbation Theory (HMChPT) \cite{Yan:1992gz,Wise:1992hn,Burdman:1992gh}. However, the predicted decay rates are, in general, unexpectedly large and not recovered in the soft meson region. Therefore, it is important to reexamine and clarify the existing calculations.  In this work we will follow Ref. \cite{Cheng:2013dua} to assume the momentum dependence of nonresonance amplitudes in the exponential form $e^{-\alpha_{\mathrm{NR}}p_B\cdot(p_i+p_j)}$ ($\alpha_\mathrm{NR}$ is unknown parameter, $p_B$, $p_i$ and $p_j$ are the four momenta of the $B$, $i$ and $j$ mesons, respectively) so that the HMChPT results are recovered in the soft meson limit $p_i,p_j\rightarrow0$. At any rate, it is important to understand and identify the underlying mechanism for nonresonant decays.

Besides the nonresonance background, the three-body meson decays are generally dominated by intermediate resonances, namely, they proceed via quasi-two-body decays containing resonance states. LHCb also observed the large $CP$ asymmetry in the localized region of the phase space \cite{Aaij:2013sfa, Aaij:2013bla}, i.e. $\mathcal{A_{CP}}(K^-\pi^+\pi^-)=0.678\pm0.078\pm0.0323\pm0.007$ for $m_{K^-\pi^+}^2<15$ $\mathrm{GeV}^2$ and $0.08<m_{\pi^+\pi^-}^2<0.66$ $\mathrm{GeV}^2$, which spans the $[\pi\pi]$ channel and $[K\pi]$ channel resonances, such as $\sigma(600)$, $\rho^0(770)$, $\omega(782)$, $K^*(892)$, $K^*(1410)$, $K_0^*(1430)$, $K^*(1680)$ and $K_2^*(1430)$ mesons. Some other considerations also motivate a precise analysis of $B^-\rightarrow K^-\pi^+\pi^-$ decays. The CP asymmetries in the decays $B\rightarrow K^*(892)\pi$, $B\rightarrow K^*(1430)\pi$ and
$B\rightarrow K_2^*(1430)\pi$ are predicted to be negligible \cite{Beneke:2003zv,Chiang:2003pm} compared to the current precision, since these are mediated by $b\rightarrow s$ loop (penguin) transitions only, with no
$b\rightarrow u$ tree component. It is worthwhile to study the contributions from $K\pi$ channel resonances in the $B^-\rightarrow K^-\pi^+\pi^-$ decays.

Theoretically, to calculate the hadronic matrix elements of hadronic $B$ weak decays, some approaches, including QCD factorization (QCDF) \cite{Beneke:2003zv,Beneke:2001ev}, perturbative QCD(pQCD) \cite{Keum:2000ph} and soft-collinear effective theory (SCET) \cite{Bauer:2000ew}, have been fully developed and extensively employed in recent years. Even though the annihilation contributions are formally power suppressed in the heavy quark limit, they may be numerically important for realistic hadronic $B$ decays, particularly for pure annihilation processes and direct $CP$ asymmetries. Unfortunately, in the collinear factorization approximation, the calculation of annihilation corrections always suffers from end-point divergence. In the pQCD approach, such divergence is regulated by introducing the parton transverse momentum $k_T$ and the Sudakov factor at the expense of modeling the additional $k_T$ dependence of meson wave functions, and large complex annihilation corrections are presented \cite{Lu:2000em}. In the SCET approach, such divergence is removed by separating the physics at different momentum scales and using zero-bin subtraction to avoid double counting the soft degrees of freedom \cite{Manohar:2006nz,Arnesen:2006vb}. In the QCDF approach, such divergence is usually parameterized in a model-independent manner \cite{Beneke:2003zv,Beneke:2001ev} and will be explicitly expressed in Sect. ${\mathrm{\uppercase\expandafter{\romannumeral3}}}$.

 There are many experimental studies which have been successfully carried out at $B$ factories (BABAR and Belle), Tevatron (CDF and D0) and LHCb and are being continued at LHCb and Belle experiments. These experiments provide highly fertile ground for theoretical studies and have yielded many exciting and important results, such as measurements of pure annihilation $B_s\rightarrow \pi \pi$ and $B_d\rightarrow K K$ decays reported recently by CDF, LHCb and Belle \cite{Aaltonen:2011jv,Aaij:2012as,Duh:2012ie}, which may suggest the existence of unexpected large annihilation contributions and have attracted much attention \cite{Xiao:2011tx,Gronau:2012gs,Chang:2014rla}. So it is also important to consider the annihilation contributions to $B$ decays.

The remainder of this paper is organized as follows. In Sect. ${\mathrm{\uppercase\expandafter{\romannumeral2}}}$, we present the form factors, decay constants and distribution amplitudes of different mesons. In Sect. ${\mathrm{\uppercase\expandafter{\romannumeral3}}}$, we present the formalism for $B$ decays in the QCDF approach. In Sect. ${\mathrm{\uppercase\expandafter{\romannumeral4}}}$, we present detailed calculations of $CP$ violation for $B^-\rightarrow K^-\pi^+\pi^-$ and $B^-\rightarrow K^-\sigma(600)$ decays. The numerical results are given in Sect. ${\mathrm{\uppercase\expandafter{\romannumeral5}}}$ and we summarize our work in Sect. ${\mathrm{\uppercase\expandafter{\romannumeral6}}}$.

\section{FORM FACTORS, DECAY CONSTANTS AND LIGHT-CONE DISTRIBUTION AMPLITUDES}

Since the form factors for $B\rightarrow P$, $B\rightarrow V$, $B\rightarrow S$ and $B\rightarrow T$ ($P$, $V$, $S$ and $T$ represent pseudoscalar, vector, scalar and tensor mesons, respectively) weak transitions and light-cone distribution amplitudes and decay constants of $P$, $V$, $S$ and $T$ will be used in treating $B$ decays, we first discuss them in this section.

The form factors of $B$ to a meson weak transition can be decomposed as \cite{Wirbel:1985ji,Cheng:2010yd}
\begin{equation}\label{bv}
\begin{split}
\langle P(p')|\hat{V}_\mu|B(p)\rangle&=\bigg(p_\mu-\frac{m_B^2-m_P^2}{q^2}q_\mu\bigg) F_1^{BP}(q^2)+\frac{m_B^2-m_P^2}{q^2}q_\mu F_0^{BP}(q^2),\\
\langle V(p')|\hat{V}_\mu|B(p)\rangle&=\frac{2}{m_B+m_V}\varepsilon_{\mu\nu\rho\sigma} \epsilon^{*\nu} p^\rho p'^\sigma V^{BV}(q^2),\\
\langle V(p')|\hat{A}_\mu|B(p)\rangle&=i\bigg\{(m_B+m_V)\epsilon_\mu^*A_1^{BV}(q^2)-\frac{\epsilon^*\cdot q}{m_B+m_V}P_\mu  A_2^{BV}(q^2)\\
&-2m_V\frac{\epsilon^*\cdot P}{q^2} q_\mu [A_3^{BV}(q^2)- A_0^{BV}(q^2)]\bigg\},\\
\langle S(p')|\hat{A}_\mu|B(p)\rangle&=-i\bigg[\bigg(P_\mu-\frac{m_B^2-m_S^2}{q^2}q_\mu\bigg)F_1^{BS}(q^2)+\frac{m_B^2-m_S^2}{q^2}q_\mu F_0^{BS}(q^2)\bigg],\\
\langle T(p')|\hat{V}_\mu|B(p)\rangle&=\frac{2}{m_B+m_T}\varepsilon_{\mu\nu\rho\sigma} e^{*\nu} p^\rho p'^\sigma V^{BT}(q^2),\\
\langle T(p')|\hat{A}_\mu|B(p)\rangle&=i\bigg\{(m_B+m_T)e_\mu^*A_1^{BT}(q^2)-\frac{e^*\cdot q}{m_B+m_T}P_\mu  A_2^{BT}(q^2)\\
&-2m_T\frac{e^*\cdot P}{q^2} q_\mu [A_3^{BT}(q^2)- A_0^{BT}(q^2)]\bigg\},\\
 \end{split}
\end{equation}
where $P_\mu=(p+p')_\mu$, $q_\mu=(p-p')_\mu$, $\hat{V}_\mu$, $\hat{A}_\mu$ and $\hat{S}_\mu$ are the weak vector, axial-vector and scalar currents, respectively, i.e. $\hat{V}_\mu=\bar{q}_2\gamma_\mu q_1, \hat{A}_\mu=\bar{q}_2\gamma_\mu \gamma_5q_1, \hat{S}=\bar{q}_2 q_1$, $\epsilon_\mu$ is the polarization vector of $V$, $e^{*\mu}\equiv \epsilon^{*\mu\nu}p_\nu/m_B$ ($\epsilon_{\mu\nu}$ is the polarization tensor of $T$), $F_i^{BP}(q^2)$ $(i=0,1)$ and $A_i^{BV(T)}(q^2)$ $(i=0,1,2,3)$ are the weak form factors. The form factors included in our calculations satisfy $F_1^{BP}(0)=F_0^{BP}(0)$, $A_3^{BV(T)}(0)=A_0^{BV(T)}(0)$, $A_3^{BV(T)}(q^2)=[(m_B+m_{V(T)})/(2m_{V(T)})]A_1^{BV(T)}(q^2)-[(m_B+m_{V(T)})/(2m_{V(T)})]A_2^{BV(T)}(q^2)$ and $F_1^{BS}(q^2)=F_0^{BS}(q^2)$.

 The decay constants are defined as \cite{Cheng:2010yd}
\begin{equation}\label{dc}
\begin{split}
\langle P(p')|\hat{A}_\mu|0\rangle&=-if_P p'_\mu,\\
\langle V(p')|\hat{V}_\mu|0\rangle&=f_Vm_V\epsilon^*_\mu, \quad\langle V(p')|\overline{q}\sigma_{\mu\nu}q'|0\rangle=f_V^\perp (p'_\mu\epsilon^*_\nu-p'_\nu\epsilon^*_\mu)m_V,\\
\langle S(p')|\hat{V}_\mu|0\rangle&=f_S p'_\mu, \quad \langle S(p')|\hat{S}|0\rangle=m_S\bar{f}_S,\\
\langle T(p')|J_{\mu\nu}(0)|0\rangle&=f_Tm_T^2\epsilon^*_{\mu\nu},\quad \langle T(p')|J_{\mu\nu\alpha}^\perp(0)|0\rangle=-if_T^\perp (p'_\nu\epsilon^*_{\mu\alpha}-p'_\mu\epsilon^*_{\mu\alpha})m_T,\\
\end{split}
\end{equation}
where $J_{\mu\nu}(0)$ and $J_{\mu\nu\alpha}^\perp(0)$ are local currents involving covariant derivatives which take the following forms:
\begin{equation}\label{J}
\begin{split}
J_{\mu\nu}(0)&=\frac{1}{2}(\bar{q}_1(0)\gamma_\mu i\overleftrightarrow{D}_\nu q_2(0)+\bar{q}_1(0)\gamma_\nu i\overleftrightarrow{D}_\mu q_2(0)),\\
J_{\mu\nu\alpha}^\perp(0)&=\bar{q}_1(0)\sigma_{\mu\nu}i\overleftrightarrow{D}_\alpha q_2(0),
\end{split}
\end{equation}
and $\overleftrightarrow{D}=\overrightarrow{D}_\mu-\overleftarrow{D}_\mu$ with $\overrightarrow{D}_\mu=\overrightarrow{\partial}_\mu+ig_sA_\mu^a \lambda^a/2$ and $\overleftarrow{D}_\mu=\overleftarrow{\partial}_\mu-ig_sA_\mu^a \lambda^a/2$ ($g_s$ is the QCD coupling constant, $A_\mu^a$ is the vector field and $\lambda^a$ are the Gellman matrices).

The twist-2 light-cone distribution amplitudes (LCDA) for the pseudoscalar, vector and tensor mesons are respectively \cite{Beneke:2003zv,Cheng:2010yd}
\begin{equation}\label{phiM}
\Phi_{M}(x,\mu)=6x(1-x)\bigg[\sum\limits_{m=0}^\infty \alpha_m^{M}(\mu)C_m^{3/2}(2x-1)\bigg], \quad M=P,V,T
 \end{equation}
and the twist-3 ones are respectively
 \begin{equation}\label{phim}
\begin{split}
\Phi_m(x)=
\begin{cases}
1& \quad m=p,  \\
3\bigg[2x-1+\sum\limits_{m=1}^\infty \alpha_{m,\perp}^V(\mu)P_{m+1}(2x-1)\bigg]& \quad m=v, \\
5\bigg(1-6x+6x^2\bigg),& \quad m=t,
\end{cases}
\end{split}
 \end{equation}
where $C_m^{3/2}$ and $P_m$ are the Gegenbauer and Legendre polynomials in Eq. (\ref{phiM}) and Eq. (\ref{phim}), respectively, $\alpha_m(\mu)$ are Gegenbauer moments which depend on the scale $\mu$.
 The twist-2 light-cone distribution amplitude for a scalar meson reads \cite{Cheng:2005nb,Cheng:2007st}
\begin{equation}\label{PhiS}
\Phi_S(x,\mu)^{(n,s)}=\bar{f}^{n,s}_S6x(x-1)\sum_{m=1,3,5}^\infty B_m(\mu)C_m^{3/2}(2x-1),
 \end{equation}
 where $B_m$ are Gegenbauer moments, $\bar{f}_S$ is the decay constant of the scalar mesons, $n$ denotes the $u$, $d$ quark component of the scalar meson, $n=\frac{1}{\sqrt{2}}(u\bar{u}+d\bar{d})$, and $s$ denotes the components $s\bar{s}$. As for the twist-3 ones, we shall take the asymptotic forms \cite{Cheng:2005nb,Cheng:2007st}
 \begin{equation}
 \Phi_s(x)^{(n,s)}=\bar{f}^{n,s}_S.
 \end{equation}
\section{B DECAYS IN QCD FACTORIZATION}
In the SM, the effective weak Hamiltonian for non-leptonic $B$-meson decays is given by \cite{Buchalla:1995vs}
 \begin{equation}\label{Hamiltonian}
 \mathcal{H}_{eff}=\frac{G_F}{\sqrt{2}}\bigg[\sum_{p=u,c}\sum_{D=d,s}\lambda_{p}^{(D)}(c_1O_1^p+c_2Q_2^p+\sum_{i=3}^{10}c_iO_i+c_{7\gamma}O_{7\gamma}+c_{8g}O_{8g})\bigg]+h.c.,
 \end{equation}
 where $\lambda_p^{(D)}=V_{pb}V_{pD}^*$, $V_{pb}$ and $V_{pD}$ are the CKM matrix elements, $G_F$ represents the Fermi constant, $c_i$ $(i=1-10,7\gamma,8g)$ are Wilson coefficients, $O_{1,2}^p$ are the tree level operators, $O_{3-6}$ are the QCD penguin operators, $O_{7-10}$ arise from electroweak penguin diagrams, and $O_{7\gamma}$ and$O_{8g}$ are the electromagnetic and chromomagnetic dipole operators, respectively.

Within the framework of QCD factorization \cite{Beneke:2003zv,Beneke:2001ev}, the effective Hamiltonian matrix elements are written in the form

\begin{equation}
\langle{M_1M_2}|\mathcal{H}_{eff}|B\rangle=\sum_{p=u,c}\lambda_{p}^{(D)}\langle{M_1M_2}|\mathcal{T}_A^p+\mathcal{T}_B^p|B\rangle,
\end{equation}
where $\mathcal{T}_A^p$ describes the contribution from naive factorization, vertex correction, penguin amplitude and spectator scattering expressed in terms of the parameters $a_i^p$, while $\mathcal{T}_B^p$ contains annihilation topology amplitudes characterized by the annihilation parameters $b_i^p$.

The flavor parameters $a_i^p$ are basically the Wilson coefficients in conjunction with short-distance nonfactorizable corrections such as vertex corrections and hard spectator interactions. In general, they have the expressions \cite{Beneke:2003zv}
\begin{equation}\label{a}
\begin{split}
a_i^p{(M_1M_2)}&={(c'_i+\frac{c'_{i\pm1}}{N_c})}N_i{(M_2)}+\frac{c'_{i\pm1}}{N_c}\frac{C_F\alpha_s}{4\pi}{\bigg[V_i{(M_2)}+\frac{4\pi^2}{N_c}H_i{(M_1M_2)}\bigg]+P_i^p{(M_2)}},
\end{split}
 \end{equation}
 where $c'_i$ are effective Wilson coefficients which are defined as $c_i(m_b)\langle O_i(m_b)\rangle=c'_i\langle O_i\rangle^{tree}$, with $\langle O_i\rangle^{tree}$ being the matrix element at the tree level, the upper (lower) signs apply when $i$ is odd (even), $N_i{(M_2)}$ is leading-order coefficient, $C_F={(N_c^2-1)}/{2N_c}$ with $N_c=3$, the quantities $V_i{(M_2)}$ account for one-loop vertex corrections, $H_i{(M_1M_2)}$ describe hard spectator interactions with a hard gluon exchange between the emitted meson and the spectator quark of the $B$ meson, and $P_i^p{(M_1M_2)}$ are from penguin contractions \cite{Beneke:2003zv}.

The expressions of the quantities $N_i(M_2)$ read
\begin{equation}
\begin{split}
N_i{(V)}=
\begin{cases}
0& \quad i=6,8,  \\
1& \quad \text{else}, \\
\end{cases}\quad N_i{(S)}=0,\quad N_i{(P)}=0,\quad N_i{(T)}=0.
\end{split}
 \end{equation}

 When $M_1M_2=VP,PV$, the correction from the hard gluon exchange between $M_2$ and the spectator quark is given by \cite{Beneke:2003zv,Beneke:2001ev}

 \begin{equation}\label{H1}
\begin{split}
 H_i{(M_1M_2)}&=\frac{f_Bf_{M_1}}{2m_V\epsilon_{V}^*\cdot p_BF_0^{B\rightarrow M_1}(0)}\int_{0}^{1}\frac{d\xi}{\xi}\Phi_{B}{(\xi)}\int_{0}^{1}dx\int_{0}^{1}dy{\bigg[\frac{\Phi_{M_2}{(x)}\Phi_{M_1}{(y)}}{\bar{x}\bar{y}}+{r_\chi^{M_1}}\frac{\Phi_{M_2}{(x)}\Phi_{m_1}{(y)}}{x\bar{y}}\bigg]},
\end{split}
\end{equation}
for $i=1-4,9,10$,
\begin{equation}\label{H2}
\begin{split}
 H_i{(M_1M_2)}&=-\frac{f_Bf_{M_1}}{2m_V\epsilon_{V}^*\cdot p_BF_0^{B\rightarrow M_1}(0)}\int_{0}^{1}\frac{d\xi}{\xi}\Phi_{B}{(\xi)}\int_{0}^{1}dx\int_{0}^{1}dy{\bigg[\frac{\Phi_{M_2}{(x)}\Phi_{M_1}{(y)}}{{x}\bar{y}}+{r_\chi^{M_1}}\frac{\Phi_{M_2}{(x)}\Phi_{m_1}{(y)}}{\bar{x}\bar{y}}\bigg]},
\end{split}
\end{equation}
for $i=5,7$ and $H_i(M_1M_2)=0$ for $i=6,8$.

When $M_1M_2=SP,PS$ \cite{Beneke:2003zv,Cheng:2005nb,Cheng:2007st},

\begin{equation}\label{H3}
\begin{split}
 H_i{(M_1M_2)}&=\frac{f_Bf_{M_1}}{f_{M_2}F_0^{B\rightarrow M_1}m_B^2}\int_{0}^{1}\frac{d\xi}{\xi}\Phi_{B}{(\xi)}\int_{0}^{1}dx\int_{0}^{1}dy{\bigg[\frac{\Phi_{M_2}{(x)}\Phi_{M_1}{(y)}}{\bar{x}\bar{y}}+{r_\chi^{M_1}}\frac{\Phi_{M_2}{(x)}\Phi_{m_1}{(y)}}{x\bar{y}}\bigg]},
\end{split}
\end{equation}
for $i=1-4,9,10$,
\begin{equation}\label{H4}
\begin{split}
 H_i{(M_1M_2)}&=-\frac{f_Bf_{M_1}}{f_{M_2}F_0^{B\rightarrow M_1}m_B^2}\int_{0}^{1}\frac{d\xi}{\xi}\Phi_{B}{(\xi)}\int_{0}^{1}dx\int_{0}^{1}dy{\bigg[\frac{\Phi_{M_2}{(x)}\Phi_{M_1}{(y)}}{{x}\bar{y}}+{r_\chi^{M_1}}\frac{\Phi_{M_2}{(x)}\Phi_{m_1}{(y)}}{\bar{x}\bar{y}}\bigg]},
\end{split}
\end{equation}
for $i=5,7$ and $H_i(M_1M_2)=0$ for $i=6,8$.

When $M_1M_2=TP,PT$ \cite{Cheng:2010yd,Cheng:2010hn}
\begin{equation}\label{H5}
\begin{split}
H_i{(M_1M_2)}&=\frac{f_Bf_{M_1}}{2m_Bp_c}\int_{0}^{1}\frac{d\xi}{\xi}\Phi_{B}{(\xi)}\int_{0}^{1}dx\int_{0}^{1}dy\\
&\begin{cases}
\frac{m_{M_1}}{2\sqrt{\frac{2}{3}}p_cA_0^{B\rightarrow M_1}(m_{M_2}^2)}{\bigg[\sqrt{\frac{2}{3}}\frac{\Phi_{M_2}{(x)}\Phi_{M_1}{(y)}}{\bar{x}\bar{y}}+{r_\chi^{M_1}}\frac{\Phi_{M_2}{(x)}\Phi_{m_1}{(y)}}{\sqrt{\frac{2}{3}}x\bar{y}}\bigg]},& \quad \text{($M_1M_2=TP$)}\\
\frac{1}{F_1^{B\rightarrow M_1}(m_{M_2}^2)}{\bigg[\frac{\Phi_{M_2}{(x)}\Phi_{M_1}{(y)}}{\bar{x}\bar{y}}+{r_\chi^{M_1}}\frac{\Phi_{M_2}{(x)}\Phi_{m_1}{(y)}}{x\bar{y}}\bigg]},&  \quad \text{($M_1M_2=PT$)}\\
\end{cases}\\
\end{split}
 \end{equation}
for $i=1-4,9,10$,

\begin{equation}\label{H6}
\begin{split}
H_i{(M_1M_2)}&=-\frac{f_Bf_{M_1}}{2m_Bp_c}\int_{0}^{1}\frac{d\xi}{\xi}\Phi_{B}{(\xi)}\int_{0}^{1}dx\int_{0}^{1}dy\\
&\begin{cases}
\frac{m_{M_1}}{2\sqrt{\frac{2}{3}}p_cA_0^{B\rightarrow M_1}(m_{M_2}^2)}{\bigg[\sqrt{\frac{2}{3}}\frac{\Phi_{M_2}{(x)}\Phi_{M_1}{(y)}}{x\bar{y}}+{r_\chi^{M_1}}\frac{\Phi_{M_2}{(x)}\Phi_{m_1}{(y)}}{\sqrt{\frac{2}{3}}\bar{x}\bar{y}}\bigg]},& \quad \text{($M_1M_2=TP$)}\\
\frac{1}{F_1^{B\rightarrow M_1}(m_{M_2}^2)}{\bigg[\frac{\Phi_{M_2}{(x)}\Phi_{M_1}{(y)}}{x\bar{y}}+{r_\chi^{M_1}}\frac{\Phi_{M_2}{(x)}\Phi_{m_1}{(y)}}{\bar{x}\bar{y}}\bigg]},&  \quad \text{($M_1M_2=PT$)}\\
\end{cases}\\
\end{split}
 \end{equation}
for $i=5,7$ and $H_i(M_1M_2)=0$ for $i=6,8$.

In Eqs. (\ref{H1}-\ref{H6}) $\bar{x}=1-x$, $\bar{y}=1-y$, and $r_\chi^{M_i}$ (i=1,2) are ``chirally-enhanced" terms which are defined as
\begin{equation}
\begin{split}
r_\chi^{P}(\mu)&=\frac{2m_P^2}{m_b(\mu)(m_{q_1}+m_{q_2})(\mu)},\quad r_\chi^{V,T}=\frac{2m_{V,T}}{m_b(\mu)}\frac{f_{V,T}^\bot(\mu)}{f_{V,T}},\\
r_\chi^{S}&=\frac{2m_S}{m_b(\mu)}\frac{\bar{f}_S(\mu)}{f_S}=\frac{2m_S^2}{m_b(\mu)(m_2(\mu)-m_1(\mu))},\quad \bar{r}_\chi^{S}=\frac{2m_S}{m_b(\mu)}.\\
\end{split}
\end{equation}

The weak annihilation contributions to $B\rightarrow M_1M_2$ can be described in terms of $b_i$ and $b_{i,EW}$, which have the following expressions:
\begin{equation}\label{b}
\begin{split}
&b_1=\frac{C_F}{N_c^2}c'_1A_1^i, \quad b_2=\frac{C_F}{N_c^2}c'_2A_1^i, \\
&b_3^p=\frac{C_F}{N_c^2}\bigg[c'_3A_1^i+c'_5(A_3^i+A_3^f)+N_cc'_6A_3^f \bigg],\quad b_4^p=\frac{C_F}{N_c^2}\bigg[c'_4A_1^i+c'_6A_2^i \bigg], \\
&b_{3,EW}^p=\frac{C_F}{N_c^2}\bigg[c'_9A_1^i+C'_7(A_3^i+A_3^f)+N_cc'_8A_3^f \bigg],\\
&b_{4,EW}^p=\frac{C_F}{N_c^2}\bigg[c'_{10}A_1^i+c'_8A_2^i \bigg],
\end{split}
 \end{equation}
 where the subscripts 1, 2, 3 of $A_n^{i,f}(n=1,2,3)$ stand for the annihilation amplitudes induced from $(V-A)(V-A)$, $(V-A)(V+A)$, and $(S-P)(S+P)$ operators, respectively, the superscripts $i$ and $f$ refer to gluon emission from the initial- and final-state quarks, respectively. Their explicit expressions are given by \cite{Beneke:2003zv,Cheng:2010yd,Cheng:2010hn,Cheng:2005nb,Cheng:2007st}
\begin{equation}\label{Ai}
\begin{split}
 A_1^i&=\pi\alpha_s\int_0^1 dx dy\begin{cases}
 \bigg(\Phi_{M_2}(x)\Phi_{M_1}(y)\bigg[\frac{1}{y(1-x\bar{y})}+\frac{1}{\bar{x}^2y}\bigg]-r_\chi^{M_1}r_\chi^{M_2} \Phi_{m_2}(x)\Phi_{m_1}(y)\frac{2}{\bar{x}y}\bigg),\quad \text{for $M_1M_2=VP,PS,$}\\
 \bigg(\Phi_{M_2}(x)\Phi_{M_1}(y)\bigg[\frac{1}{y(1-x\bar{y})}+\frac{1}{\bar{x}^2y}\bigg]+r_\chi^{M_1}r_\chi^{M_2} \Phi_{m_2}(x)\Phi_{m_1}(y)\frac{2}{\bar{x}y}\bigg),\quad \text{for $M_1M_2=PV,SP,$}\\
 \sqrt{\frac{2}{3}}\bigg(\Phi_{M_2}(x)\Phi_{M_1}(y)\bigg[\frac{1}{y(1-x\bar{y})}+\frac{1}{\bar{x}^2y}\bigg]-\frac{3}{2}r_\chi^{M_1}r_\chi^{M_2} \Phi_{m_2}(x)\Phi_{m_1}(y)\frac{2}{\bar{x}y}\bigg),\quad \text{for $M_1M_2=TP,$}\\
 \sqrt{\frac{2}{3}}\bigg(\Phi_{M_2}(x)\Phi_{M_1}(y)\bigg[\frac{1}{y(1-x\bar{y})}-\frac{1}{\bar{x}^2y}\bigg]+\frac{3}{2}r_\chi^{M_1}r_\chi^{M_2} \Phi_{m_2}(x)\Phi_{m_1}(y)\frac{2}{\bar{x}y}\bigg),\quad \text{for $M_1M_2=PT,$}\\
 \end{cases}\\
 A_2^i&=\pi\alpha_s\int_0^1 dx dy\begin{cases}
 \bigg(-\Phi_{M_2}(x)\Phi_{M_1}(y)\bigg[\frac{1}{\bar{x}(1-x\bar{y})}+\frac{1}{\bar{x}y^2}\bigg]+r_\chi^{M_1}r_\chi^{M_2} \Phi_{m_2}(x)\Phi_{m_1}(y)\frac{2}{\bar{x}y}\bigg),\quad \text{for $M_1M_2=VP,PS,$}\\
 \bigg(-\Phi_{M_2}(x)\Phi_{M_1}(y)\bigg[\frac{1}{\bar{x}(1-x\bar{y})}+\frac{1}{\bar{x}y^2}\bigg]-r_\chi^{M_1}r_\chi^{M_2} \Phi_{m_2}(x)\Phi_{m_1}(y)\frac{2}{\bar{x}y}\bigg),\quad \text{for $M_1M_2=PV,SP,$}\\
 \sqrt{\frac{2}{3}}\bigg(\Phi_{M_2}(x)\Phi_{M_1}(y)\bigg[\frac{1}{\bar{x}(1-x\bar{y})}+\frac{1}{\bar{x}y^2}\bigg]+\frac{3}{2}r_\chi^{M_1}r_\chi^{M_2} \Phi_{m_2}(x)\Phi_{m_1}(y)\frac{2}{\bar{x}y}\bigg),\quad \text{for $M_1M_2=TP,$}\\
 \sqrt{\frac{2}{3}}\bigg(\Phi_{M_2}(x)\Phi_{M_1}(y)\bigg[\frac{1}{\bar{x}(1-x\bar{y})}+\frac{1}{\bar{x}y^2}\bigg]-\frac{3}{2}r_\chi^{M_1}r_\chi^{M_2} \Phi_{m_2}(x)\Phi_{m_1}(y)\frac{2}{\bar{x}y}\bigg),\quad \text{for $M_1M_2=PT,$}\\
 \end{cases}\\
 A_3^i&=\pi\alpha_s\int_0^1 dx dy\begin{cases}
 \bigg(r_\chi^{M_1}\Phi_{M_2}(x)\Phi_{m_1}(y)\frac{2\overline{y}}{\overline{x}y(1-x\bar{y})}+r_\chi^{M_2} \Phi_{M_1}(y)\Phi_{m_2}(x)\frac{2x}{\overline{x}y(1-x\bar{y})}\bigg),\quad \text{for $M_1M_2=VP,PS,$}\\
 \bigg(-r_\chi^{M_1}\Phi_{M_2}(x)\Phi_{m_1}(y)\frac{2\overline{y}}{\overline{x}y(1-x\bar{y})}+r_\chi^{M_2} \Phi_{M_1}(y)\Phi_{m_2}(x)\frac{2x}{\overline{x}y(1-x\bar{y})}\bigg),\quad \text{for $M_1M_2=PV,SP,$}\\
 \sqrt{\frac{2}{3}}\bigg(\frac{3}{2}r_\chi^{M_1}\Phi_{M_2}(x)\Phi_{m_1}(y)\frac{2\overline{y}}{\overline{x}y(1-x\bar{y})}+r_\chi^{M_2} \Phi_{M_1}(y)\Phi_{m_2}(x)\frac{2x}{\overline{x}y(1-x\bar{y})}\bigg),\quad \text{for $M_1M_2=TP,PT,$}\\
 \end{cases}\\
 A_3^f&=\pi\alpha_s\int_0^1 dx dy\begin{cases}
 \bigg(r_\chi^{M_1}\Phi_{M_2}(x)\Phi_{m_1}(y)\frac{2(1+\overline{x})}{\overline{x}^2y}-r_\chi^{M_2} \Phi_{M_1}(y)\Phi_{m_2}(x)\frac{2(1+y)}{\overline{x}y^2}\bigg),\quad \text{for $M_1M_2=VP,PS,$}\\
 \bigg(-r_\chi^{M_1}\Phi_{M_2}(x)\Phi_{m_1}(y)\frac{2(1+\overline{x})}{\overline{x}^2y}-r_\chi^{M_2} \Phi_{M_1}(y)\Phi_{m_2}(x)\frac{2(1+y)}{\overline{x}y^2}\bigg),\quad \text{for $M_1M_2=PV,SP,$}\\
 \sqrt{\frac{2}{3}}\bigg(\frac{3}{2}r_\chi^{M_1}\Phi_{M_2}(x)\Phi_{m_1}(y)\frac{2(1+\overline{x})}{\overline{x}^2y}-r_\chi^{M_2} \Phi_{M_1}(y)\Phi_{m_2}(x)\frac{2(1+y)}{\overline{x}y^2}\bigg),\quad \text{for $M_1M_2=TP,PT,$}\\
 \end{cases}\\
 A_1^i&=A_2(M_1M_2)^f=0.\\
 \end{split}
  \end{equation}

When dealing with the weak annihilation contributions and the hard spectator contributions, one has to deal with the infrared endpoint singularity $X=\int_0^1 dx/(1-x)$. The treatment of this endpoint divergence is model dependent, and we follow Ref. \cite{Beneke:2003zv} to parameterize this endpoint divergence in the annihilation and hard spectator diagrams as
\begin{equation}\label{XH}
X_{A,H}^{M_1M_2}=(1+\rho^{M_1M_2}_{A,H} e^{i\phi_{A,H}^{M_1M_2}})\ln\frac{m_B}{\Lambda_h},
\end{equation}
where $\Lambda_h$ is a typical scale of order 0.5 $\mathrm{GeV}$, $\rho_{A(H)}^{M_1M_2}$ is an unknown real parameter and $\phi_{A(H)}^{M_1M_2}$ is a free strong phase in the range $[0,2\pi]$ for the annihilation (hard spectator) process. In our work, we will follow the assumption $X_H^{M_1M_2}=X_A^{M_1M_2}=X^{M_1M_2}$ for the $B\rightarrow PV(PT)$ decays \cite{Wang:2016yrm,Cheng:2009cn,Cheng:2010yd}, but for the $B\rightarrow SP$ decays, we will further assume that $X^{M_1M_2}= X^{M_2M_1}$ compared with the $B\rightarrow PV(PT)$ decays.

\section{CALCULATION OF CP VIOLATION}

\subsection{FRAMEWORK}
\subsubsection{Nonresonance background}

In the absence of resonances, the factorizable nonresonance amplitude for the $B^-\rightarrow K^-\pi^+\pi^-$ decay has the expression \cite{Cheng:2007si,Cheng:2013dua}
\begin{equation}\label{ANR}
\begin{split}
A_{NR}&=\frac{G_F}{\sqrt{2}}\sum_{p=u,c}\lambda_p^{s}\bigg[\langle\pi^+\pi^-|(\bar{u}b)_{V-A}|B^-\rangle\langle K^-|(\bar{s}u)_{V-A}|0\rangle[a_1\delta_{pu}+a_4^p+a_{10}^p-(a_6^p+a_8^p)r_\chi^K]\\
&+\langle \pi^-|\bar{d}b|B^-\rangle\langle K^-\pi^+|\bar{s}d|0\rangle(-2a_6^p+2a_8^p)\bigg].
\end{split}
\end{equation}
For the parameters $a_i$ which contain effective Wilson coefficients, we take the following values \cite{Cheng:2013dua,Cheng:2007si}:
\begin{equation}
\begin{split}
a_1&=0.99\pm0.037i,\quad a_2=0.19-0.11i,\quad a_3=-0.002+0.004i,\quad a_5=0.0054-0.005i,\\
a_4^u&=-0.03-0.02i,\quad a_4^c=-0.04-0.008i,\quad a_6^u=-0.006-0.02i,\quad a_6^c=-0.006-0.006i,\\
a_7&=0.54\times10^{-4}i,\quad a_8^u=(4.5-0.5i)\times10^{-4},\quad a_8^c=(4.4-0.3i)\times10^{-4},\quad a_9=-0.010-0.0002i,\\
a_{10}^u&=(-58.3+86.1i)\times10^{-5},\quad a_{10}^c=(-60.3+88.8i)\times10^{-5},
\end{split}
\end{equation}
For the current-induced process, the amplitude $\langle\pi^+\pi^-|(\bar{u}b)_{V-A}|B^-\rangle\langle K^-|(\bar{s}u)_{V-A}|0\rangle$ can be expressed in terms of three unknown form factors \cite{Lee:1992ih,Cheng:2013dua,Cheng:2007si}
\begin{equation}
\begin{split}
A_{\mathrm{current-ind}}^{\mathrm{HMChPT}}&\equiv\langle\pi^+(p_1)\pi^-(p_2)|(\bar{u}b)_{V-A}|B^-\rangle\langle K^-(p_3)|(\bar{s}u)_{V-A}|0\rangle\\
&=-\frac{f_\pi}{2}[2m_3^2r+(m_B^2-s_{12}-m_3^2)\omega_++(s_{23}-s_{13}-m_2^2+m_1^2)\omega_-],\\
\end{split}
\end{equation}
 where $r$, $\omega_{\pm}$, and $h$ are form factors which can be evaluated in the framework of HMChPT and the results read \cite{Lee:1992ih,Fajfer:1998yc}
\begin{equation}
\begin{split}
\omega_+&=-\frac{g}{f_\pi^2}\frac{f_{B^*}m_{B^*}\sqrt{m_Bm_{B^*}}}{s_{23}-m_{B^*}^2}\bigg[1-\frac{(p_B-p_1)\cdot p_1}{m_{B^*}^2}\bigg]+\frac{f_B}{2f_\pi^2},\\
\omega_-&=\frac{g}{f_\pi^2}\frac{f_{B^*}m_{B^*}\sqrt{m_Bm_{B^*}}}{s_{23}-m_{B^*}^2}\bigg[1+\frac{(p_B-p_1)\cdot p_1}{m_{B^*}^2}\bigg],\\
r&=\frac{f_B}{2f_\pi^2}-\frac{f_B}{f_\pi^2}\frac{p_B\cdot(p_2-p_1)}{(p_B-p_1-p_2)^2-m_{B^2}}\\
&+\frac{2gf_{B^*}}{f_\pi^2}\sqrt{\frac{m_B}{m_{B^*}}}\frac{(p_B-p_1)\cdot p_1}{s_{23}-m_{B^*}^2}-\frac{4g^2f_B}{f_\pi^2}\frac{m_Bm_{B^*}}{(p_B-p_1-p_2)^2-m_B^2}\\
&\times\frac{p_1\cdotp_2-p_1\cdot(p_B-p_1) p_2\cdot(p_B-p_1)/m_{B^*}^2}{s_{23}-m_{B^*}^2},
\end{split}
\end{equation}
where $s_{ij}\equiv(p_i+p_j)^2$, $g$ is a heavy-flavor-independent strong coupling which can be extracted from the CLEO measurement of the $D^{*+}$ decay width, $|g|=0.59\pm0.01\pm0.07$ \cite{Ahmed:2001xc}, which sign is fixed to be negative in Ref. \cite{Yan:1992gz}.

However, the predicted nonresonance results based on HMChPT are not recovered in the soft meson region and lead to decay rates that are too large which are in disagreement with experiment \cite{Cheng:2002qu}. For example, the branching fraction is found to be of order $7.5\times10^{-5}$, which is one order of magnitude larger than the BaBar result, $5.3\times10^{-6}$ \cite{Aubert:2009av}. The issue is related to the applicability HMChPT, which requires the two mesons in the final state in the $B\rightarrow M_1M_2$ transition have to be soft and hence an exponential form of the amplitudes is necessary \cite{Cheng:2007si,Cheng:2016ajl},
\begin{equation}
A_{\mathrm{current-ind}}=A_{\mathrm{current-ind}}^{\mathrm{HMChPT}}e^{-\alpha_\mathrm{NR}p_B\cdot(p_1+p_2)}e^{i\mathrm{\phi_{12}}},
\end{equation}
where $\alpha_{\mathrm{NR}}$ is constrained from the tree dominated decay $B^-\rightarrow \pi^+\pi^-\pi^-$ to be $\alpha_{\mathrm{NR}}=0.081_{-0.009}^{+0.015}\mathrm{GeV}^{-2}$, and the phase $\phi_{12}$ of the nonresonant amplitude will be set to zero for simplicity \cite{Cheng:2007si,Cheng:2016ajl}.

The matrix element of $\langle K^-\pi^+|\bar{s}d|0\rangle^{\mathrm{NR}}$ is related to $\langle K^+K^-|\bar{s}s|0\rangle^{\mathrm{NR}}$ via SU(3) symmetry, i.e. $\langle K^-\pi^+|\bar{s}d|0\rangle^{\mathrm{NR}}=\langle K^+K^-|\bar{s}s|0\rangle^{\mathrm{NR}}$, we shall adopt Ref. \cite{Cheng:2013dua} to assume that final state interactions amount to giving a large strong phase $\delta$ to the nonresonance component of the matrix element of $\langle K^-\pi^+|\bar{s}d|0\rangle^{\mathrm{NR}}$ and a fit to the data of direct $CP$ asymmetries in $B^-\rightarrow K^-\pi^+\pi^-$ yields
\begin{equation}\label{FF1}
\begin{split}
\langle K^-(p_1)\pi^+(p_2)|\bar{s}d|0\rangle^{\mathrm{NR}}&=\frac{\nu}{3}(3F_{\mathrm{NR}}+2F'_{\mathrm{NR}})+\sigma_{\mathrm{NR}}e^{-\alpha s_{12}}e^{i\delta}\\
&\approx\frac{\nu}{3}(3F_{\mathrm{NR}}+2F'_{\mathrm{NR}})+\sigma_{\mathrm{NR}}e^{-\alpha s_{12}}e^{i\pi}\bigg(1+4\frac{m_K^2-m_\pi^2}{s_{12}}\bigg),\\
\end{split}
\end{equation}
where the parameter $\sigma_{\mathrm{NR}}=(3.39^{+0.18}_{-0.21})e^{i\pi/4}\mathrm{GeV}$, and $\nu=\frac{m_{K^+}^2}{m_u+m_s}=\frac{m_K^2-m_\pi^2}{m_s-m_d}$ characterizes the quark-operator parameter $\langle\bar{q}q\rangle$ which spontaneously breaks the chiral symmetry and the experimental measurement leads to $\alpha=(0.14\pm0.02)\mathrm{GeV}^{-2}$ \cite{Aubert:2007sd}. Motivated by the asymptotic constraints from pQCD, namely, $F(t)^{(')}\rightarrow(1/t)[\ln(t/\tilde{\Lambda}^2)]^{-1}$ in the large-t limit \cite{Brodsky:1974vy}, the nonresonance form factors in Eq. (\ref{FF1}) can be parameterized as \cite{Cheng:2013dua}
\begin{equation}
\begin{split}\label{FNR}
F_{NR}(s_{23})&=\bigg(\frac{x_1}{s_{23}}+\frac{x_2}{s^2_{23}}\bigg)\Big[\ln\bigg(\frac{s_{23}}{\tilde{\Lambda}^2}\bigg)\Big]^{-1},\\
F'_{NR}(s_{23})&=\bigg(\frac{x'_1}{s_{23}}+\frac{x'_2}{s^2_{23}}\bigg)\Big[\ln\bigg(\frac{s_{23}}{\tilde{\Lambda}^2}\bigg)\Big]^{-1},\\
\end{split}
\end{equation}
where $\tilde{\Lambda}\approx 0.3$ $\mathrm{GeV}$ is the QCD scale parameter, the unknown parameters $x_i$ and $x'_i$ are fitted from the kaon electromagentic data, giving the following best-fit values \cite{Chua:2002pi}:
\begin{equation}\label{x}
\begin{split}
x_1&=-3.26 \mathrm{GeV}^2,\quad x_2=5.02 \mathrm{GeV}^4,\\
x'_1&=0.47 \mathrm{GeV}^2,\quad x'_2=0.\\
\end{split}
\end{equation}

\subsubsection{Resonance contributions}
LHCb has observed large $CP$ asymmetries in localized regions of phase space $m_{K^-\pi^+}^2<15$ $\mathrm{GeV}^2$ and $0.08<m_{\pi^+\pi^-}^2<0.66$ $\mathrm{GeV}^2$ \cite{Aaij:2013sfa, Aaij:2013bla}, which contains the $[\pi\pi]$ and $[K\pi]$ channel resonances including $\sigma(600)$, $\rho^0(770)$, $\omega(782)$, $K_0^*(1430)$, $K_2^*(1430)$ and $(K^*)^i$ ($K^*(892)$, $K^*(1410)$, $K^*(1680)$ for $i=1,2,3$) which will be denoted as $\sigma$, $\rho$, $\omega$, $K_0^*$, $K_2^*$ and $(K^*)^i$ for simplicity, respectively. The total resonance amplitude including the $\rho-\omega$ mixing effect can be written as \cite{Cheng:2013dua,Dedonder:2010fg}
\begin{equation}\label{AR}
\begin{split}
\sum_RA_R&=A_{\sigma}+A_{\rho,\omega}+\sum_iA_{(K^*)^i} +A_{K_0^*}+A_{K_2^*}\\
&=A_{[\pi\pi]}+A_{[K\pi]},
 \end{split}
  \end{equation}
where the sum over $R$ refers to that over the aforementioned resonances including the $\rho-\omega$ mixing effect.

$\rho-\omega$ mixing has the dual advantages that the strong phase difference is large and well known \cite{Gardner:1997yx,Guo:1998eg}. In order to deal with the large localized $CP$ violation, we need to appeal this mechanisms to the $B^-\rightarrow K^-\pi^+\pi^-$ decay. In this scenario one has \cite{Wang:2015ula,Guo:2000uc,Bediaga:2006jk}
\begin{equation}\label{Av}
\begin{split}
A_{\rho,\omega}&=\langle K^-\pi^+\pi^-|\mathcal{H}^T|B^-\rangle+\langle K^-\pi^+\pi^-|\mathcal{H}^P|B^-\rangle\\
&=\epsilon^\lambda\cdot(p_{\pi^-}-p_{\pi^+})\bigg[\bigg(\frac{g_\rho}{s_\rho s_\omega}\tilde{\Pi}_{\rho\omega}t_\omega+\frac{g_\rho}{s_\rho}t_\rho\bigg)+\bigg(\frac{g_\rho}{s_\rho s_\omega}\tilde{\Pi}_{\rho\omega}p_\omega+\frac{g_\rho}{s_\rho}p_\rho\bigg)\bigg],
  \end{split}
  \end{equation}
where $\mathcal{H}^T$ and $\mathcal{H}^P$ are the Hamiltonians for the tree and penguin operators, respectively,  $t_V$($V=\rho$ or $\omega$) is the tree amplitude and $p_V$ is the penguin amplitude for producing an intermediate vector meson $V$, $g_\rho$ is the coupling for $\rho\rightarrow\pi^+\pi^-$, $\tilde{\Pi}_{\rho\omega}$ is the effective $\rho-\omega$ mixing amplitude, and $s_V$ is from the inverse propagator of the vector meson $V$, $s_V=s-m_V^2+im_V\Gamma_V$ and $\sqrt{s}$ is the invariant mass of the $\pi^+\pi^-$ pair. The direct coupling $\omega\rightarrow\pi^+\pi^-$ is effectively absorbed into $\tilde{\Pi}_{\rho\omega}$ \cite{OConnell:1997ggd}, leading to the explicit $s$ dependence of $\tilde{\Pi}_{\rho\omega}$. Making the expansion $\tilde{\Pi}_{\rho\omega}(s)=\tilde{\Pi}_{\rho\omega}(m_\omega^2)+(s-m_\omega^2)\tilde{\Pi}_{\rho\omega}'(m_\omega^2)$, the $\rho-\omega$ mixing parameters were determined in the fit of Gardner and O'Connell \cite{Gardner:1997ie}: $\mathfrak{Re}\tilde{\Pi}_{\rho\omega}(m_\omega^2)=-3500\pm300 \mathrm{MeV^2}$, $\mathfrak{Im}\tilde{\Pi}_{\rho\omega}(m_\omega^2)=-300\pm300 \mathrm{MeV^2}$, $\tilde{\Pi}_{\rho\omega}'(m_\omega^2)=0.03\pm0.04$. In practice, the effect of the derivative term is negligible.

Because of its large width, $\sigma$ can not be modeled by a naive Breit-Wigner distribution. In this paper, we will adopt the Bugg model to parameterize the distribution of  $\sigma$ which is given by \cite{Bugg:2006gc,Aaij:2015sqa,Li:2015tja}
\begin{equation}\label{T1}
\begin{split}
 R_{\sigma}(s)=M\Gamma_1(s)/\bigg[M^2-s-g_1^2(s)\frac{s-s_A}{M^2-s_A}z(s)-iM\Gamma_{\mathrm{tot}}(s)\bigg],
 \end{split}
 \end{equation}
 where $z(s)=j_1(s)-j_1(M^2)$ with $j_1(s)=\frac{1}{\pi}[2+\rho_1\ln(\frac{1-\rho_1}{1+\rho_1})]$, $\Gamma_{\mathrm{tot}}(s)=\sum\limits_{i=1}^4 \Gamma_i(s)$ and
 \begin{equation}\label{T2}
\begin{split}
M\Gamma_1(s)&=g_1^2(s)\frac{s-s_A}{M^2-s_A}\rho_1(s),\\
M\Gamma_2(s)&=0.6g_1^2(s)(s/M^2)\mathrm{exp}(-\alpha|s-4m_K^2|)\rho_2(s),\\
M\Gamma_3(s)&=0.2g_1^2(s)(s/M^2)\mathrm{exp}(-\alpha|s-4m_\eta^2|)\rho_3(s),\\
M\Gamma_4(s)&=Mg_{4\pi}\rho_{4\pi}(s)/\rho_{4\pi}(M^2),\\
g_1^2(s)&=M(c_1+c_2s)\mathrm{exp}[-(s-M^2)/A],\\
\rho_{4\pi}(s)&=1.0/[1+\mathrm{exp}(7.082-2.845s)].\\
 \end{split}
 \end{equation}
The parameters in Eqs. (\ref{T1}, \ref{T2}) are fixed to be $M=0.953 \mathrm{GeV}$, $s_A=0.14m_\pi^2$, $c_1=1.302\mathrm{GeV}^2$,
$c_2=0.340$, $A=2.426\mathrm{GeV}^2$ and $g_{4\pi}=0.011\mathrm{GeV}$, which are given in the fourth column of Table I in Ref. \cite{Bugg:2006gc}. The parameters $\rho_{1,2,3}$ are the phase-space factors of the decay channels $\pi\pi$, $KK$ and $\eta\eta$, respectively, which are defined as \cite{Bugg:2006gc}
\begin{equation}\label{rho}
\rho_i(s)=\sqrt{1-4\frac{m_i^2}{s}},
\end{equation}
with $m_1=m_\pi$, $m_2=m_K$ and $m_3=m_\eta$. Other resonants in Eq. (\ref{AR}) will be modeled by the naive Breit-Wigner distribution.

Within the QCDF, we derive the tree and penguin amplitudes of $\rho$ and $\omega$ in Eq.(\ref{Av}) and obtain
\begin{equation}\label{trho}
\begin{split}
t_\rho=-iG_F m_\rho\epsilon_\rho^*\cdot p_B\lambda_u^{(s)}\bigg[\alpha_1(\rho K )A_0^{B\rightarrow \rho}(0)f_K+\alpha_2(K \rho)F_0^{B\rightarrow K}(0)f_\rho+b_2(\rho K)f_Bf_\rho f_K\bigg],\\
 \end{split}
 \end{equation}

 \begin{equation}\label{tomega}
\begin{split}
t_\omega&=-iG_Fm_\omega\epsilon_\omega^*\cdot p_B\lambda_u^{(s)}\bigg[\alpha_1(\omega K )A_0^{B\rightarrow \omega}(0)f_K+\alpha_2(K \omega)F_0^{B\rightarrow K}(0)f_\omega+b_2(\omega K)f_Bf_\omega f_K\bigg],\\
 \end{split}
 \end{equation}

\begin{equation}\label{prho}
\begin{split}
p_\rho&=-iG_Fm_\rho\epsilon_\rho^*\cdot p_B\sum_{p=u,c}\lambda_p^{(s)}\bigg\{\bigg[\alpha_4^p(\rho K)+\alpha_{4,EW}^p(\rho K)\bigg]A_0^{B\rightarrow \rho}(0)f_K+\frac{3}{2}\alpha_{3,EW}^p(K \rho)F_0^{B\rightarrow K}(0)f_\rho\\
&+\bigg[b_3^p(\rho K)-b_{3,EW}^p(\rho K)\bigg]f_Bf_\rho f_K\bigg\},\\
\end{split}
\end{equation}

\begin{equation}\label{pomega}
\begin{split}
p_\omega&=-iG_Fm_\omega\epsilon_\omega^*\cdot p_B\sum_{p=u,c}\lambda_p^{(s)}\bigg\{\bigg[2\alpha_3(K \omega)+\frac{1}{2}\alpha_3^p(K \omega)\bigg]F_0^{B\rightarrow K}(0)f_\omega+\bigg[\alpha_4^p(\omega K)+\frac{3}{2}\alpha_{4,EW}^p(\omega K)\bigg]\\
&\times A_0^{B\rightarrow \omega}(0)f_K+\bigg[b_3^p(\omega K)+b_{3,EW}^p(\omega K)\bigg]f_Bf_\omega f_K\bigg\}.\\
 \end{split}
 \end{equation}

 The polarization vectors of a vector meson $V$ with mass $m_V$ and momentum $p$ satisfies
\begin{equation}\label{pp}
\sum\limits_{\lambda=0,\pm1}\epsilon_\mu^\lambda(p)(\epsilon_\nu^\lambda(p))^*=-\bigg(g_{\mu\nu}-\frac{p_\mu p_\nu}{m_V^2}\bigg),
\end{equation}
from which one obtains \cite{Zhang:2013oqa}
\begin{equation}\label{p2p3}
\sum\limits_{\lambda=0,\pm1}\epsilon^\lambda\cdot(p_2-p_3)(\epsilon^{\lambda})^*\cdot p_B=\hat{s}_{13}-s_{(13)},
\end{equation}
$\hat{s}_{13}$ is the midpoint of the allowed range of $s_{13}$, i.e. $\hat{s}_{13}=(s_{13, \textrm{max}}+s_{13,\textrm{min}})/2$, with $s_{13,\textrm{max}}$ and $s_{13,\textrm{min}}$ being the maximum and minimum values of $s_{13}$ for fixed $s_{12}$.

As for the  polarization vectors of a tensor meson we have \cite{Dedonder:2010fg}
\begin{equation}\label{P2P3}
\sum\limits_{-2}^2\epsilon_{\alpha\beta}(\lambda)p^\alpha_2p^\beta_3\epsilon^*_{\mu\nu}(\lambda)p^\nu_Bp_1^\mu=\frac{1}{3}(|\vec{p}_1||\vec{p}_2|)^2-(\vec{p}_1\cdot\vec{p}_2)^2,
\end{equation}
where $\vec{p}_1$ and $\bar{p}_2$ are three momenta of $\pi^-(p_1)$ and $\pi^+(p_2)$, respectively, in the rest frame of $\pi^+(p_2)$ and $K^-(p_3)$.  One obtains, with $m_{23}=\sqrt{s_{23}}$ \cite{Dedonder:2010fg},
\begin{equation}\label{p1p2}
\begin{split}
|\vec{p}_1|&=\frac{1}{2m_{23}}\sqrt{[m_B^2-(m_{23}+m_1)^2][m_B^2-(m_{23}-m_1)^2]},\\
|\vec{p}_2|&=\frac{1}{2m_{23}}\sqrt{[s_{23}-(m_3+m_2)^2][s_{23}-(m_3-m_2)^2]},\\
\vec{p}_1\cdot\vec{p}_2&=s_{12}-s_{23}+\frac{(m_B^2-m_1^2)(m_3^2-m_2^2)}{s_{23}}.\\
  \end{split}
  \end{equation}
Inserting Eqs. (\ref{trho}-\ref{pomega}) into Eq. (\ref{Av}), one can get the amplitude from $\rho-\omega$ mixing contribution
\begin{equation}\label{Arhoomega}
\begin{split}
A_{\rho,\omega}&=-iG_F\bigg(\hat{s}_{K\pi}-s_{K\pi}\bigg)\bigg\{\frac{g_\rho}{s_\rho s_\omega}\tilde{\Pi}_{\rho\omega}\bigg[m_\omega\lambda_u^{(s)}\bigg(\alpha_1(\omega K )A_0^{B\rightarrow \omega}(0)f_K
+\alpha_2(K \omega)F_0^{B\rightarrow K}(0)f_\omega\\
&+b_2(\omega K)f_Bf_\omega f_Km_\omega/(m_Bp_c)\bigg)\bigg]+\frac{g_\rho}{s_\rho}\bigg[m_\rho\lambda_u^{(s)}\bigg(\alpha_1(\rho K )A_0^{B\rightarrow \rho}(0)f_K+\alpha_2(K \rho)F_0^{B\rightarrow K}(0)f_\rho\\
&+b_2(\rho K)f_Bf_\rho f_Km_\omega/(m_Bp_c)\bigg]\bigg\}
+\bigg\{\frac{g_\rho}{s_\rho s_\omega}\tilde{\Pi}_{\rho\omega}\times\bigg[m_\omega\sum_{p=u,c}\lambda_p^{(s)}\bigg\{\bigg(2\alpha_3(K \omega)+\frac{1}{2}\alpha_3^p(K \omega)\bigg)F_0^{B\rightarrow K}(0)f_\omega\\
&+\bigg(\alpha_4^p(\omega K)+\frac{3}{2}\alpha_{4,EW}^p(\omega K)\bigg)A_0^{B\rightarrow \omega}(0)f_K+\bigg(b_3^p(\omega K)+b_{3,EW}^p(\omega K)\bigg)
 f_Bf_\omega f_Km_\omega/(m_Bp_c)\bigg\}\bigg]\\
 &+\frac{g_\rho}{s_\rho}\bigg[m_\rho\sum_{p=u,c}\lambda_p^{(s)}\bigg\{\bigg(\alpha_4^p(\rho K)+\alpha_{4,EW}^p(\rho K)\bigg) A_0^{B\rightarrow \rho}(0)f_K\\
 &+\frac{3}{2}\alpha_{3,EW}^p(K \rho)F_0^{B\rightarrow K}(0)f_\rho+\bigg(b_3^p(\rho K)-b_{3,EW}^p(\rho K)\bigg)f_Bf_\rho f_Km_\omega/(m_Bp_c)\bigg\}\bigg]\bigg\},\\
  \end{split}
  \end{equation}
 where $p_c$ is the magnitude of the three momentum of either final state meson in the rest frame of the $B$ meson, $\alpha_i^p(M_1M_2)$ can be expressed in terms of the cofficients $a_i^p$ defined in Eq. (\ref{a}) and have the following expressions:
 \begin{equation}\label{HSP}
 \begin{split}
 \alpha_1{(M_1M_2)}&=a_1{(M_1M_2)},\\
\alpha_2{(M_1M_2)}&=a_2{(M_1M_2)},\\
\alpha_3^p{(M_1M_2)}&=\begin{cases}
a_3^p{(M_1M_2)}-a_5^p{(M_1M_2)}, \quad \text{if $M_1M_2=VP,SP,TP$},\\
a_3^p{(M_1M_2)}+a_5^p{(M_1M_2)}, \quad \text{if $M_1M_2=PV,PS,PT$},\\
\end{cases}\\
\alpha_4^p{(M_1M_2)}&=\begin{cases}
a_4^p{(M_1M_2)}+r_\chi^{M_2}a_6^p{(M_1M_2)}, \quad \text{if $M_1M_2=PV,PT$},\\
a_4^p{(M_1M_2)}-r_\chi^{M_2}a_6^p{(M_1M_2)},\quad \text{if $M_1M_2=VP,PS,SP,TP$},\\
\end{cases}\\
\alpha_{3,EW}^p{(M_1M_2)}&=\begin{cases}
a_9^p{(M_1M_2)}-a_7^p{(M_1M_2)},\quad \text{if $M_1M_2=VP,SP,TP$},\\
a_9^p{(M_1M_2)}+a_7^p{(M_1M_2)},\quad \text{if $M_1M_2=PV,PS,PT$},\\
\end{cases}\\
\alpha_{4,EW}^p{(M_1M_2)}&=\begin{cases}
 a_{10}^p{(M_1M_2)}+r_\chi^{M_2}a_8^p{(M_1M_2)},\quad \text{if $M_1M_2=PV,PT$},\\
 a_{10}^p{(M_1M_2)}-r_\chi^{M_2}a_8^p{(M_1M_2)},\quad \text{if $M_1M_2=VP,PS,SP,TP$}.\\
 \end{cases}\\
  \end{split}
  \end{equation}

Meanwhile, it is straightforward get the amplitudes contributed by others resonances, including $\sigma$, $(K^*)^i$, $K^*_0$ and $K^*_2$, respectively,
\begin{equation}\label{Asigma}
\begin{split}
A_\sigma &=iG_Fg_{\sigma\pi\pi}R_\sigma\sum_{p=u,c}\lambda_p^{(s)}
  \bigg\{(m_{\sigma}^2-m_B^2)F_0^{B\rightarrow f}(m_K^2)
 f_K\bigg[\delta_{pu}\alpha_1(\sigma K)+\alpha_4^p(\sigma K)+\alpha_{4,EW}^p(\sigma K)\bigg]\\
 &-f_Bf_K\bar{f}^u_{\sigma}\bigg[\delta_{pu}b_2(\sigma K)+b_3^p(\sigma K)+b_{3,EW}^p(\sigma K)\bigg]
   +\bigg[\delta_{pu}\alpha_2(K \sigma)+2\alpha_3^p(K\sigma)+\frac{1}{2}\alpha_{3,EW}^p(K\sigma)\bigg]\\
   &\times(m_B^2-m_K^2)F_0^{B\rightarrow K}(0)\bar{f}^u_{\sigma}
  +\bigg[\sqrt{2}\alpha_3^p(K \sigma)+\sqrt{2}\alpha_4^p(K \sigma)-\frac{1}{\sqrt{2}}\alpha_{3,EW}^p(K \sigma)-\frac{1}{\sqrt{2}}\alpha_{4,EW}^p(K \sigma)\bigg]\\
  &\times(m_B^2-m_K^2)F_0^{B\rightarrow K}(m_\sigma^2)\bar{f}^s_{\sigma}
  -f_Bf_K\bar{f}^s_{\sigma}\bigg[\sqrt{2}\delta_{pu}b_2(K\sigma)+\sqrt{2}b_3^p(K\sigma)+\sqrt{2}b_{3,EW}^p(K\sigma)\bigg]\bigg\},\\
\end{split}
\end{equation}

\begin{equation}\label{AK}
\begin{split}
A_{(K^*)^i}&=-iG_F\bigg(\hat{s}_{\pi\pi}-s_{\pi\pi}\bigg)\frac{g_{{(K^*)^i}K\pi}}{s_{V}}\sum_{p=u,c}\lambda_p^{(s)}\bigg\{b_2(\pi {(K^*)^i})f_Bf_\pi f_{(K^*)^i}m_{{(K^*)^i}}/(m_Bp_c)\\
&-\bigg(\alpha_4^p(\pi {(K^*)^i})-\frac{1}{2}\alpha_{4,EW}^p(\pi {(K^*)^i})\bigg)\bigg(-2m_VF_1^{B\rightarrow \pi}f_{(K^*)^i}\bigg)
-\bigg(b_3^p(\pi {(K^*)^i})+b_{3,EW}^p(\pi {(K^*)^i})\bigg)\\
&\times f_Bf_\pi f_{(K^*)^i}m_{(K^*)^i}/(m_Bp_c)\bigg\},\\
\end{split}
\end{equation}
where $(K^*)^i=K^*(892),K^*(1410),K^*(1680)$ corresponding to $i=1,2,3$, respectively, and

\begin{equation}\label{AK0}
\begin{split}
A_{K_0^*}&=-iG_F\frac{g_{K^*_0K\pi}}{s_{K^*_0}}\sum_{p=u,c}\lambda_p^{(s)}\bigg\{b_2(\pi K_0^*)f_Bf_\pi \bar{f}_{K_0^*}-\bigg(\alpha_4^p(\pi K_0^*)-\frac{1}{2}\alpha_{4,EW}^p(\pi K_0^*)\bigg)\\
&\times\bigg((m_B^2-m_\pi^2)F_0^{B\rightarrow \pi}(m_{K_0^*}^2)\bar{f}_{K_0^*}\bigg)
-\bigg(b_3^p(\pi K_0^*)+b_{3,EW}^p(\pi K_0^*)\bigg)f_Bf_\pi \bar{f}_{K_0^*}\bigg\}.\\
\end{split}
\end{equation}

\begin{equation}\label{AK2}
\begin{split}
A_{K_2^*}&=-iG_F\bigg[\frac{1}{3}\bigg(|\vec{p}_{\pi^-}||\vec{p}_{\pi^+}|\bigg)^2-\bigg(\vec{p}_{\pi^-}\cdot\vec{p}_{\pi^+}\bigg)^2\bigg]\frac{g_{K_2^*K\pi}}{s_{K_2^*}}\sum_{p=u,c}\lambda_p^{(s)}\bigg\{b_2(\pi K_2^*)f_Bf_\pi f_{K_2^*}m_{K_2^*}/(m_Bp_c)\\
&-\bigg(\alpha_4^p(\pi K_2^*)-\frac{1}{2}\alpha_{4,EW}^p(\pi K_2^*)\bigg)\bigg(-2m_TF_1^{B\rightarrow \pi}f_{K_2^*}\bigg)
-\bigg(b_3^p(\pi K_2^*)+b_{3,EW}^p(\pi K_2^*)\bigg)f_Bf_\pi f_{K_2^*}m_{K_2^*}/(m_Bp_c)\bigg\}.\\
\end{split}
\end{equation}

Combining Eq. (\ref{Arhoomega}) with Eq. (\ref{Asigma}), one obtain the amplitude of $B^-\rightarrow K^-[\pi^+\pi^-]\rightarrow K^-\pi^+\pi^-$
\begin{equation}\label{Apipi}
\begin{split}
A_{[\pi\pi]}&=-iG_F\bigg(\hat{s}_{K\pi}-s_{K\pi}\bigg)\bigg\{\frac{g_\rho}{s_\rho s_\omega}\tilde{\Pi}_{\rho\omega}\bigg[m_\omega\lambda_u^{(s)}\bigg(\alpha_1(\omega K )A_0^{B\rightarrow \omega}(0)f_K
+\alpha_2(K \omega)F_0^{B\rightarrow K}(0)f_\omega\\
&+b_2(\omega K)f_Bf_\omega f_Km_\omega/(m_Bp_c)\bigg)\bigg]+\frac{g_\rho}{s_\rho}\bigg[m_\rho\lambda_u^{(s)}\bigg(\alpha_1(\rho K )A_0^{B\rightarrow \rho}(0)f_K+\alpha_2(K \rho)F_0^{B\rightarrow K}(0)f_\rho\\
&+b_2(\rho K)f_Bf_\rho f_Km_\omega/(m_Bp_c)\bigg]\bigg\}
+\bigg\{\frac{g_\rho}{s_\rho s_\omega}\tilde{\Pi}_{\rho\omega}\times\bigg[m_\omega\sum_{p=u,c}\lambda_p^{(s)}\bigg\{\bigg(2\alpha_3(K \omega)+\frac{1}{2}\alpha_3^p(K \omega)\bigg)F_0^{B\rightarrow K}(0)f_\omega\\
&+\bigg(\alpha_4^p(\omega K)+\frac{3}{2}\alpha_{4,EW}^p(\omega K)\bigg)A_0^{B\rightarrow \omega}(0)f_K+\bigg(b_3^p(\omega K)+b_{3,EW}^p(\omega K)\bigg)
 f_Bf_\omega f_Km_\omega/(m_Bp_c)\bigg\}\bigg]\\
 &+\frac{g_\rho}{s_\rho}\bigg[m_\rho\sum_{p=u,c}\lambda_p^{(s)}\bigg\{\bigg(\alpha_4^p(\rho K)+\alpha_{4,EW}^p(\rho K)\bigg) A_0^{B\rightarrow \rho}(0)f_K+\frac{3}{2}\alpha_{3,EW}^p(K \rho)F_0^{B\rightarrow K}(0)f_\rho\\
 &+\bigg(b_3^p(\rho K)-b_{3,EW}^p(\rho K)\bigg)f_Bf_\rho f_Km_\omega/(m_Bp_c)\bigg\}\bigg]\bigg\}+iG_Fg_{\sigma\pi\pi}R_\sigma\sum_{p=u,c}\lambda_p^{(s)}\bigg\{(m_{\sigma}^2-m_B^2)
 F_0^{B\rightarrow f}(m_K^2)f_K\\
 &\times\bigg[\delta_{pu}\alpha_1(\sigma K)+\alpha_4^p(\sigma K)+\alpha_{4,EW}^p(\sigma K)\bigg]-f_Bf_K\bar{f}^u_{\sigma}\bigg[\delta_{pu}b_2(\sigma K)+b_3^p(\sigma K)+b_{3,EW}^p(\sigma K)\bigg]\\
   &+\bigg[\delta_{pu}\alpha_2(K \sigma)+2\alpha_3^p(K\sigma)+\frac{1}{2}\alpha_{3,EW}^p(K\sigma)\bigg]\times(m_B^2-m_K^2)F_0^{B\rightarrow K}(0)\bar{f}^u_{\sigma}
  +\bigg[\sqrt{2}\alpha_3^p(K \sigma)+\sqrt{2}\alpha_4^p(K \sigma)\\
  &-\frac{1}{\sqrt{2}}\alpha_{3,EW}^p(K \sigma)-\frac{1}{\sqrt{2}}\alpha_{4,EW}^p(K \sigma)\bigg](m_B^2-m_K^2)F_0^{B\rightarrow K}(m_\sigma^2)\bar{f}^s_{\sigma}\\
  &-f_Bf_K\bar{f}^s_{\sigma}\bigg[\sqrt{2}\delta_{pu}b_2(K\sigma)+\sqrt{2}b_3^p(K\sigma)+\sqrt{2}b_{3,EW}^p(K\sigma)\bigg]\bigg\},\\
\end{split}
\end{equation}
Meanwhile, using the Eqs. (\ref{AK}-\ref{AK2}), we get the amplitude of $B^-\rightarrow [K^-\pi^+]\pi^-\rightarrow K^-\pi^+\pi^-$
\begin{equation}\label{AKpi}
\begin{split}
A_{[K\pi]}&=-iG_F\bigg(\hat{s}_{\pi\pi}-s_{\pi\pi}\bigg)\frac{g_{{(K^*)^i}K\pi}}{s_{V}}\sum_{p=u,c}\lambda_p^{(s)}\bigg\{b_2(\pi {(K^*)^i})f_Bf_\pi f_{(K^*)^i}m_{{(K^*)^i}}/(m_Bp_c)\\
&-\bigg(\alpha_4^p(\pi {(K^*)^i})-\frac{1}{2}\alpha_{4,EW}^p(\pi {(K^*)^i})\bigg)\bigg(-2m_VF_1^{B\rightarrow \pi}f_{(K^*)^i}\bigg)
-\bigg(b_3^p(\pi {(K^*)^i})+b_{3,EW}^p(\pi {(K^*)^i})\bigg)\\
&\times f_Bf_\pi f_{(K^*)^i}m_{(K^*)^i}/(m_Bp_c)\bigg\}-iG_F\frac{g_{K^*_0K\pi}}{s_{K^*_0}}\sum_{p=u,c}\lambda_p^{(s)}\bigg\{b_2(\pi K_0^*)f_Bf_\pi \bar{f}_{K_0^*}-\bigg(\alpha_4^p(\pi K_0^*)-\frac{1}{2}\alpha_{4,EW}^p(\pi K_0^*)\bigg)\\
&\times\bigg((m_B^2-m_\pi^2)F_0^{B\rightarrow \pi}(m_{K_0^*}^2)\bar{f}_{K_0^*}\bigg)
-\bigg(b_3^p(\pi K_0^*)+b_{3,EW}^p(\pi K_0^*)\bigg)f_Bf_\pi \bar{f}_{K_0^*}\bigg\}
-iG_F\bigg[\frac{1}{3}\bigg(|\vec{p}_{\pi^-}||\vec{p}_{\pi^+}|\bigg)^2\\
&-\bigg(\vec{p}_{\pi^-}\cdot\vec{p}_{\pi^+}\bigg)^2\bigg]\frac{g_{K_2^*K\pi}}{s_{K_2^*}}\sum_{p=u,c}\lambda_p^{(s)}\bigg\{b_2(\pi K_2^*)f_Bf_\pi f_{K_2^*}m_{K_2^*}/(m_Bp_c)-\bigg(\alpha_4^p(\pi K_2^*)-\frac{1}{2}\alpha_{4,EW}^p(\pi K_2^*)\bigg)\\
&\times(-2m_TF_1^{B\rightarrow \pi}f_{K_2^*})-\bigg(b_3^p(\pi K_2^*)+b_{3,EW}^p(\pi K_2^*)\bigg)f_Bf_\pi f_{K_2^*}m_{K_2^*}/(m_Bp_c)\bigg\}.\\
\end{split}
\end{equation}

\subsubsection{Total result for the amplitude of $B^-\rightarrow K^-\pi^+\pi^-$}
In the QCDF, both the resonance and nonresonance contributions have been considered, inserting Eqs. (\ref{Apipi}) and (\ref{AKpi}) to Eq. (\ref{AR}) then combing the Eqs. (\ref{ANR}-\ref{x}), the decay amplitude via $B^-\rightarrow R+NR\rightarrow K^-\pi^+\pi^-$ can be finally obtained as:
\begin{equation}\label{AA}
\begin{split}
A&=iG_Fg_{\sigma\pi\pi}R_\sigma\sum_{p=u,c}\lambda_p^{(s)}
  \bigg\{(m_{\sigma}^2-m_B^2)F_0^{B\rightarrow f}(m_K^2)
 f_K\bigg[\delta_{pu}\alpha_1(\sigma K)+\alpha_4^p(\sigma K)+\alpha_{4,EW}^p(\sigma K)\bigg]\\
 &-f_Bf_K\bar{f}^u_{\sigma}\bigg[\delta_{pu}b_2(\sigma K)+b_3^p(\sigma K)+b_{3,EW}^p(\sigma K)\bigg]
   +\bigg[\delta_{pu}\alpha_2(K \sigma)+2\alpha_3^p(K\sigma)+\frac{1}{2}\alpha_{3,EW}^p(K\sigma)\bigg]\\
   &\times(m_B^2-m_K^2)F_0^{B\rightarrow K}(0)\bar{f}^u_{\sigma}
  +\bigg[\sqrt{2}\alpha_3^p(K \sigma)+\sqrt{2}\alpha_4^p(K \sigma)-\frac{1}{\sqrt{2}}\alpha_{3,EW}^p(K \sigma)-\frac{1}{\sqrt{2}}\alpha_{4,EW}^p(K \sigma)\bigg]\\
  &\times(m_B^2-m_K^2)F_0^{B\rightarrow K}(m_\sigma^2)\bar{f}^s_{\sigma}
  -f_Bf_K\bar{f}^s_{\sigma}\bigg[\sqrt{2}\delta_{pu}b_2(K\sigma)+\sqrt{2}b_3^p(K\sigma)+\sqrt{2}b_{3,EW}^p(K\sigma)\bigg]\bigg\}\\
  &-iG_F\bigg(\hat{s}_{K\pi}-s_{K\pi}\bigg)\bigg\{\frac{g_\rho}{s_\rho s_\omega}\tilde{\Pi}_{\rho\omega}\bigg[m_\omega\lambda_u^{(s)}\bigg(\alpha_1(\omega K )A_0^{B\rightarrow \omega}(0)f_K
+\alpha_2(K \omega)F_0^{B\rightarrow K}(0)f_\omega\\
&+b_2(\omega K)f_Bf_\omega f_Km_\omega/(m_Bp_c)\bigg)\bigg]+\frac{g_\rho}{s_\rho}\bigg[m_\rho\lambda_u^{(s)}\bigg(\alpha_1(\rho K )A_0^{B\rightarrow \rho}(0)f_K+\alpha_2(K \rho)F_0^{B\rightarrow K}(0)f_\rho\\
&+b_2(\rho K)f_Bf_\rho f_Km_\omega/(m_Bp_c)\bigg]\bigg\}
+\bigg\{\frac{g_\rho}{s_\rho s_\omega}\tilde{\Pi}_{\rho\omega}\times\bigg[m_\omega\sum_{p=u,c}\lambda_p^{(s)}\bigg\{\bigg(2\alpha_3(K \omega)+\frac{1}{2}\alpha_3^p(K \omega)\bigg)F_0^{B\rightarrow K}(0)f_\omega\\
&+\bigg(\alpha_4^p(\omega K)+\frac{3}{2}\alpha_{4,EW}^p(\omega K)\bigg)A_0^{B\rightarrow \omega}(0)f_K+\bigg(b_3^p(\omega K)+b_{3,EW}^p(\omega K)\bigg)
 f_Bf_\omega f_Km_\omega/(m_Bp_c)\bigg\}\bigg]\\
 &+\frac{g_\rho}{s_\rho}\bigg[m_\rho\sum_{p=u,c}\lambda_p^{(s)}\bigg\{\bigg(\alpha_4^p(\rho K)+\alpha_{4,EW}^p(\rho K)\bigg) A_0^{B\rightarrow \rho}(0)f_K
+\frac{3}{2}\alpha_{3,EW}^p(K \rho)F_0^{B\rightarrow K}(0)f_\rho\\
&+\bigg(b_3^p(\rho K)-b_{3,EW}^p(\rho K)\bigg)f_Bf_\rho f_Km_\omega/(m_Bp_c)\bigg\}\bigg]\bigg\}
-iG_F\frac{g_{{(K^*)^i}K\pi}}{s_{V}}\sum_{p=u,c}\lambda_p^{(s)}\bigg\{b_2(\pi {(K^*)^i})f_Bf_\pi f_{{(K^*)^i}}\\
&-\bigg(\alpha_4^p(\pi {(K^*)^i})-\frac{1}{2}\alpha_{4,EW}^p(\pi {(K^*)^i})\bigg)\bigg(-2m_Bp_cF_1^{B\rightarrow \pi}f_{(K^*)^i}\bigg)
-\bigg(b_3^p(\pi {(K^*)^i})+b_{3,EW}^p(\pi {(K^*)^i})\bigg)f_Bf_\pi f_{(K^*)^i}\bigg\}\\
&-iG_F\bigg(\hat{s}_{\pi\pi}-s_{\pi\pi}\bigg)\frac{g_{K^*_0K\pi}}{s_{K^*_0}}\sum_{p=u,c}\lambda_p^{(s)}\bigg\{b_2(\pi K_0^*)f_Bf_\pi \bar{f}_{K_0^*}-\bigg(\alpha_4^p(\pi K_0^*)-\frac{1}{2}\alpha_{4,EW}^p(\pi K_0^*)\bigg)\\
&\times\bigg((m_B^2-m_\pi^2)F_0^{B\rightarrow \pi}(m_{K_0^*}^2)\bar{f}_{K_0^*}\bigg)
-\bigg(b_3^p(\pi K_0^*)+b_{3,EW}^p(\pi K_0^*)\bigg)f_Bf_\pi \bar{f}_{K_0^*}\bigg\}\\
&-iG_F\bigg[\frac{1}{3}\bigg(|\vec{p}_{\pi^-}||\vec{p}_{\pi^+}|\bigg)^2-\bigg(\vec{p}_{\pi^-}\cdot\vec{p}_{\pi^+}\bigg)^2\bigg]\frac{g_{K_2^*K\pi}}{s_{K_2^*}}\sum_{p=u,c}\lambda_p^{(s)}\bigg\{b_2(\pi K_2^*)f_Bf_\pi f_{K_2^*}m_{K_2^*}/(m_Bp_c)\\
&-\bigg(\alpha_4^p(\pi K_2^*)-\frac{1}{2}\alpha_{4,EW}^p(\pi K_2^*)\bigg)\bigg(-2m_TF_1^{B\rightarrow \pi}f_{K_2^*}\bigg)
-\bigg(b_3^p(\pi K_2^*)+b_{3,EW}^p(\pi K_2^*)\bigg)f_Bf_\pi f_{K_2^*}m_{K_2^*}/(m_Bp_c)\bigg\}\\
&-\frac{G_F}{\sqrt{2}}\sum_{p=u,c}\lambda_p^{s}\frac{f_\pi}{2}\bigg[2m_K^2r+\bigg(m_B^2-s_{\pi\pi}-m_K^2\bigg)\omega_++\bigg(s_{\pi K}-s_{\pi K}\bigg)\omega_-\bigg]
\bigg[a_1\delta_{pu}+a_4^p+a_{10}^p-\bigg(a_6^p+a_8^p\bigg)r_\chi^K\bigg]\\
&\times e^{-\alpha_\mathrm{NR}(s_{\pi\pi}+s_{K\pi}-m_\pi^2-m_K^2)}
+\bigg(\frac{m_B^2-m_\pi^2}{m_d-m_b}F_0^{B\rightarrow\pi}(0)\bigg)\bigg(-2a_6^p+2a_8^p\bigg)\\
&\times\bigg[\frac{\nu}{3}(3F_{NR}+2F'_{NR})+\sigma_{NR}e^{-\alpha s_{\pi\pi}}e^{i\pi}\bigg(1+4\frac{m_K^2-m_\pi^2}{s_{\pi\pi}}\bigg)\bigg].\\
\end{split}
\end{equation}

\subsubsection{Localizd CP violation}
Totally, the decay amplitude for $B\rightarrow K^-\pi^+\pi^-$ is the sum of resonant (R) contributions and the nonresonant (NR) background \cite{Cheng:2013dua}
\begin{equation}\label{A}
A=\sum_R A_R+A_{NR}.
 \end{equation}
 The differential CP asymmetry parameter can be defined as
 \begin{equation}\label{CP asymmetry parameter}
\mathcal{A_{CP}}=\frac{|A|^2-|\bar{A}|^2}{|A|^2+|\bar{A}|^2}.
 \end{equation}

In this work, we will consider eight resonances in a certain phase region $\Omega$ which includes $m_{K^-\pi^+}^2<15$ $\mathrm{GeV}^2$ and $0.08<m_{\pi^+\pi^-}^2<0.66$ $\mathrm{GeV}^2$ for the $B^{-}\rightarrow K^-\pi^+\pi^-$ decay. By integrating the denominator and numerator of $\mathcal{A_{CP}}$ in this region, we get the localized integrated $CP$ asymmetry, which can be measured by experiments and takes the following form:
  \begin{equation}\label{localized CP}
\mathcal{A^\mathrm{\Omega}_{CP}}=\frac{\int_\Omega ds_{12}ds_{13}(|A|^2-|\bar{A}|^2)}{\int_\Omega ds_{12}ds_{13}(|A|^2+|\bar{A}|^2)}.
 \end{equation}
\subsection{Calculation of differential $CP$ violation and branching fraction of $B^-\rightarrow K^- \sigma$ decay}
Using Eq. (\ref{CP asymmetry parameter}), the differential $CP$ asymmetry parameter of $B\rightarrow M_1M_2$ can be expressed as
 \begin{equation}\label{CP asymmetry}
\mathcal{A_{CP}}(B\rightarrow M_1M_2)=\frac{|A(B\rightarrow M_1M_2)|^2-|\bar{A}(B\rightarrow M_1M_2)|^2}{|A(B\rightarrow M_1M_2)|^2+|\bar{A}(B\rightarrow M_1M_2)|^2}.
 \end{equation}
The branching fraction of the $B\rightarrow M_1M_2$ decay has the following form:
\begin{equation}\label{e1}
\mathcal{B}(B\rightarrow M_1M_2)=\tau_B \frac{p_c}{8\pi m_B^2}|A(B\rightarrow M_1M_2)|^2,
\end{equation}
where $\tau_B$ and $m_B$ are the lifetime and the mass of the $B$ meson, respectively, $p_c$ is the magnitude of the three momentum of either final state meson in the rest frame of the $B$ meson which can be expressed as
\begin{equation}\label{e2}
p_c=\frac{1}{2m_B}\sqrt{[m_B^2-(m_{M_1}+m_{M_2})^2][m_B^2-(m_{M_1}-m_{M_2})^2]},
\end{equation}
with $m_{M_1}$ and $m_{M_2}$ being the two final state mesons' masses, respectively.

 The amplitude of $B^-\rightarrow K^- \sigma$ has the following form:
\begin{equation}\label{abk}
\begin{split}
A(B^-\rightarrow \sigma K^-)&=\langle \sigma K^-|\mathcal{H}_{eff}|B^-\rangle\\
&=\sum_{p=u,c}\lambda_p^{(s)}\frac{G_F}{2}\bigg\{\bigg[\alpha_1(\sigma K)\delta_{pu}+\alpha_4^p(\sigma K)+
\alpha_{4,EW}^p(\sigma K)\bigg]\times\bigg(m_{\sigma}^2-m_B^2\bigg)F_0^{B\rightarrow \sigma}(m_K^2)f_K\\
&+\bigg[\alpha_2(K \sigma)\delta_{pu}+2\alpha_{3}(K \sigma)+\frac{1}{2}\alpha_{3,EW}^p(K \sigma))\bigg]\times (m_B^2-m_{K}^2)F_0^{B\rightarrow K}(m_{\sigma}^2)\bar{f}^u_{\sigma}\\
&+\bigg[\sqrt{2}\alpha_3^p(K \sigma)+\sqrt{2}\alpha_{4}^p(K \sigma)-\frac{1}{\sqrt{2}}\alpha_{3,EW}^p(K \sigma)-\frac{1}{\sqrt{2}}\alpha_{4,EW}^p(K \sigma)\bigg]\\
&\times \bigg(m_B^2-m_{K}^2\bigg)F_0^{B\rightarrow K}(m_{\sigma}^2)\bar{f}^s_{\sigma}-\bigg[b_2(\sigma K)\delta_{pu}+b_3^p(\sigma K)+b_{3,EW}^p(\sigma K)\bigg]\\
&\times f_Bf_K\bar{f}^u_{\sigma}-\sqrt{2}\bigg[b_2(K \sigma)\delta_{pu}+b_3^p(K \sigma)+b_{3,EW}^p(K \sigma)\bigg]\times f_Bf_K\bar{f}^s_{\sigma}\bigg\}.
\end{split}
\end{equation}
Substituting Eq. (\ref{abk}) into Eq. (\ref{CP asymmetry}) we can get the expression of $\mathcal{A_{CP}}(B^-\rightarrow K^- \sigma)$. Substituting Eqs. (\ref{abk}) and (\ref{e2}) into Eq. (\ref{e1}), one can obtain the branching fraction of $B^-\rightarrow K^- \sigma$.

\begin{figure}[ht]
\begin{minipage}{0.4\linewidth}
\centerline{\includegraphics[width=1\textwidth]{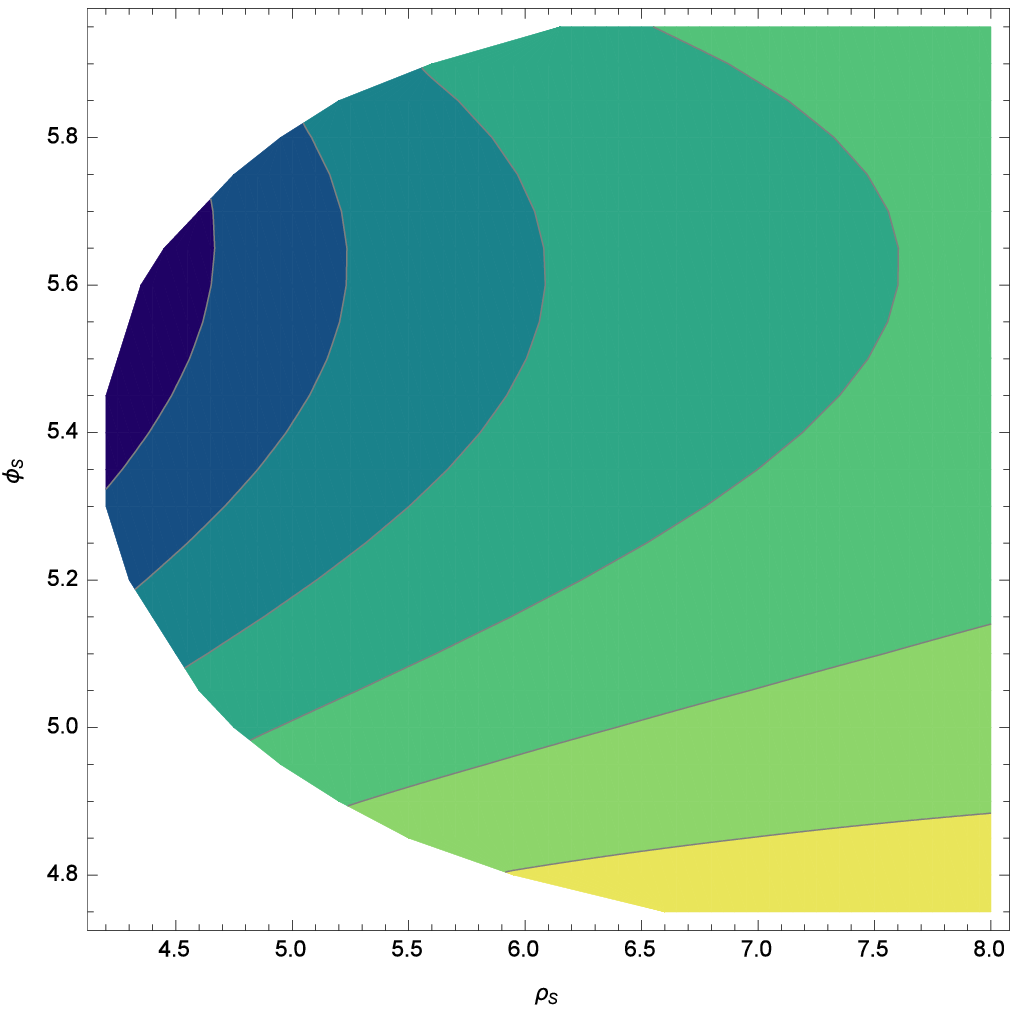}}
\centerline{}
\end{minipage}
\qquad
\begin{minipage}{0.07\linewidth}
\centerline{\includegraphics[width=0.8\textwidth]{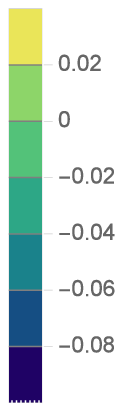}}
\centerline{}
\end{minipage}
\caption{Numerical results of $\mathcal{A_{CP}}(B^{-}\rightarrow K^-\sigma)$ as functions of $\rho_S$ and $\phi_S$.}
\label{p1}
\end{figure}

\begin{figure}[ht]
\begin{minipage}{0.4\linewidth}
\centerline{\includegraphics[width=1\textwidth]{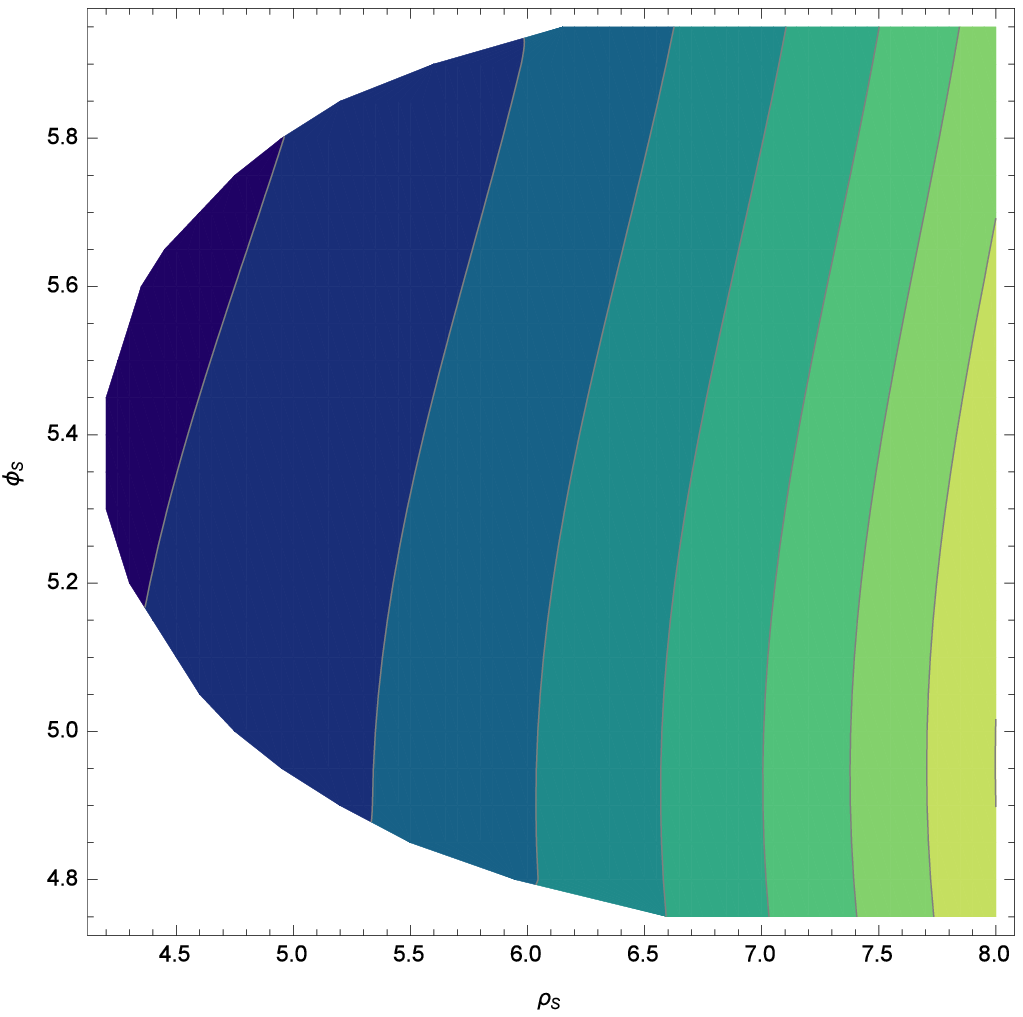}}
\centerline{}
\end{minipage}
\qquad
\begin{minipage}{0.06\linewidth}
\centerline{\includegraphics[width=0.8\textwidth]{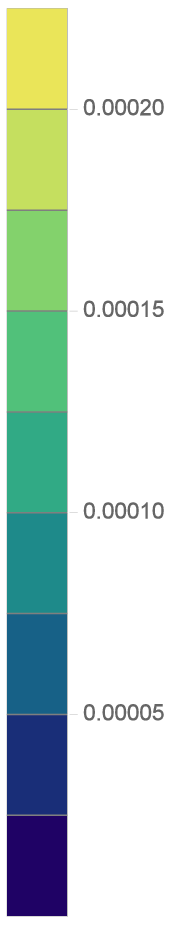}}
\centerline{}
\end{minipage}
\caption{Numerical results of $\mathcal{B}(B^-\rightarrow K^-\sigma)$ $(\times10^5)$ as functions of $\rho_S$ and $\phi_S$.}
\label{p2}
\end{figure}

\section{Numerical results}
The theoretical results obtained in the QCDF approach depend on many input parameters. The values of the Wolfenstein parameters are given as $\bar{\rho}=0.117\pm0.021$, $\bar{\eta}=0.353\pm0.013$ \cite{Agashe:2014kda}.

The effective Wilson coefficients used in our calculations are taken from Ref. \cite{Wang:2014hba}:
\begin{equation}\label{C}
\begin{split}
&C'_1=-0.3125, \quad C'_2=-1.1502, \\
&C'_3=2.120\times10^{-2}+5.174\times10^{-3}i,\quad C'_4=-4.869\times10^{-2}-1.552\times10^{-2}i, \\
&C'_5=1.420\times10^{-2}+5.174\times10^{-3}i,\quad C'_6=-5.792\times10^{-2}-1.552\times10^{-2}i, \\
&C'_7=-8.340\times10^{-5}-9.938\times10^{-5}i,\quad C'_8=3.839\times10^{-4}, \\
&C'_9=-1.017\times10^{-2}-9.938\times10^{-5}i,\quad C'_{10}=1.959\times10^{-3}. \\
\end{split}
\end{equation}
For the masses appeared in $B$ decays, we shall use the following values (in units of $\mathrm{GeV}$) \cite{Agashe:2014kda}:
\begin{equation}
\begin{split}
m_u&=m_d=0.0035,\quad m_s=0.119, \quad m_b=4.2,\quad m_q=\frac{m_u+m_d}{2},\quad m_{\pi^\pm}=0.14,\\
m_{B^-}&=5.279,\quad m_\omega=0.782,\quad m_{\rho^0(770)}=0.775,\quad m_{K^-}=0.494,\quad m_{K^*}(892)=0.895,\\
m_{K^*}(1410)&=1.414,\quad m_{K^*_0}(1430)=1.425,\quad m_{K^*}(1680)=1.717,\quad m_{K^*_2}(1430)=1.426,\\
\end{split}
\end{equation}
while for the widths we shall use (in units of $\mathrm{GeV}$) \cite{Agashe:2014kda}
\begin{equation}
\begin{split}
\Gamma_{\rho}&=0.149,\quad\Gamma_{\omega}=0.00849,\quad\Gamma_{\sigma(600)}=0.5,\quad\Gamma_{K^*(892)}=0.047,\quad \Gamma_{K^*(1410)}=0.232,\\
\Gamma_{K^*(1680)}&=0.322,\quad \Gamma_{K^*_0(1430)}=0.270,\quad \Gamma_{K^*_2(1430)}=0.109,\\
\Gamma_{\rho\rightarrow\pi\pi}&=0.149,\quad\Gamma_{\omega\rightarrow\pi\pi}=0.00013,\quad\Gamma_{\sigma(600)\rightarrow\pi\pi}=0.3,\quad\Gamma_{K^*(892)\rightarrow K\pi}=0.0487,\\
\Gamma_{K^*(1410)\rightarrow K\pi}&=0.015,\quad \Gamma_{K^*(1680)\rightarrow K\pi}=0.10,\quad \Gamma_{K^*_0(1430)\rightarrow K\pi}=0.251,\quad \Gamma_{K^*_2(1430)\rightarrow K\pi}=0.054.\\
\end{split}
\end{equation}
The strong coupling constants are determined from the measured widths through the relations \cite{Cheng:2013dua,Dedonder:2010fg,Dedonder:2014xpa}
\begin{equation}\label{gSVT}
\begin{split}
g_{S\rightarrow M'_1M'_2}=\sqrt{\frac{8\pi m_S^2}{p_c(S)^2}\Gamma_{S\rightarrow M'_1M'_2}},\\
g_{V\rightarrow M'_1M'_2}=\sqrt{\frac{6\pi m_V^2}{p_c(V)^3}\Gamma_{V\rightarrow M'_1M'_2}},\\
g_{T\rightarrow M'_1M'_2}=\sqrt{\frac{60\pi m_T^2}{p_c(T)^5}\Gamma_{T\rightarrow M'_1M'_2}},\\
\end{split}
\end{equation}
where $p_c(S,V,T)$ are the magnitudes of the three momenta of the final state meson in the rest frame of $S$, $V$, and $T$ mesons, respectively.

The following numerical values for the decay constants will be used (in units of $\mathrm{GeV}$)\cite{Cheng:2013dua,Cheng:2010yd,Cheng:2005nb,Qi:2018wkj}:
\begin{equation}
\begin{split}
f_{\pi^\pm}&=0.131,\quad f_{B^-}=0.21\pm0.02, \quad f_{K^-}=0.156\pm0.007, \quad \bar{f}_\sigma^u=0.4829\pm0.14, \quad \bar{f}^s_\sigma=-0.21\pm0.10,\\
 f_{\rho^0(770)}&=0.216\pm0.003,\quad f_{\rho^0(770)}^\perp =0.165\pm0.009,\quad f_\omega=0.187\pm0.005,\quad f_\omega^\perp=0.151\pm0.009,\\
 f_{K^*(892)}&=0.22\pm0.005,\quad f_{K^*(892)}^\perp=0.185\pm0.010, \quad f_{K^*(1410)}=0.22\pm0.1,\\
 f^\perp_{K^*(1410)}&=0.185\pm0.1, \quad f_{K^*(1680)}=0.22\pm0.005,\quad f_{K^*(1680)}^\perp=0.185\pm0.010,\\
\bar{f}_{K^*_0(1430)}&=0.445\pm0.050,\quad f_{K_2^*(1430)}=0.118\pm0.005,\quad f_{K_2^*(1430)}^\perp=0.077\pm0.014.\\
\end{split}
\end{equation}
As for the form factors, we use \cite{Cheng:2013dua,Cheng:2010yd,Cheng:2005nb}
\begin{equation}
\begin{split}
F_0^{B\rightarrow K}(0)&=0.35\pm0.04,\quad F_0^{B\rightarrow \sigma}(m_K^2)=0.45\pm0.15,\quad A_0^{B\rightarrow \rho}(0)=0.303\pm0.029,\\
A_0^{B\rightarrow K^*(892)}(0)&=0.374\pm0.034,\quad A_0^{B\rightarrow K^*(1410)}(0)=0.4\pm0.1,\quad A_0^{B\rightarrow K^*(1680)}(0)=0.4\pm0.1,\\
A_0^{B\rightarrow K_2^*(1430)}(0)&=0.25\pm0.04,\quad A_1^{B\rightarrow K_2^*(1430)}(0)=0.14\pm0.02,\quad F_0^{B\rightarrow \pi}(0)=0.25\pm0.03, \\
F_0^{B\rightarrow K_0^*(1430) }(0)&=0.26.\\
\end{split}
\end{equation}
The values of Gegenbauer moments at $\mu=1 \mathrm{GeV}$ are taken from \cite{Cheng:2013dua,Cheng:2010yd,Cheng:2005nb,Qi:2018wkj}:
\begin{equation}
\begin{split}
\alpha_1^\rho&=0,\quad \alpha_2^\rho=0.15\pm0.07, \quad \alpha_{1,\perp}^\rho=0,\quad \alpha_{2,\perp}^\rho=0.14\pm0.06, \\
 \alpha_1^\omega&=0,\quad \alpha_2^\omega=0.15\pm0.07, \quad \alpha_{1,\perp}^\omega=0,\quad \alpha_{2,\perp}^\omega=0.14\pm0.06, \\
\alpha_1^{K^*_2(1430)}&=\frac{5}{3},\quad \alpha_{1,\perp}^{K^*_2(1430)}=\frac{5}{3},\\
\alpha_1^{K^*(892)}&=0.03\pm0.02,\quad \alpha_{1,\perp}^{K^*(892)}=0.04\pm0.03,\quad \alpha_2^{K^*(892)}=0.11\pm0.09,\quad \alpha_{2,\perp}^{K^*(892)}=0.10\pm0.08,\\
\alpha_1^{K^*(1410)}&=0.03\pm0.1,\quad \alpha_{1,\perp}^{K^*(1410)}=0.04\pm0.1,\quad \alpha_2^{K^*(1410)}=0.11\pm0.1,\quad \alpha_{2,\perp}^{K^*(1410)}=0.10\pm0.1,\\
\alpha_1^{K^*(1680)}&=0.03\pm0.1,\quad \alpha_{1,\perp}^{K^*(1680)}=0.04\pm0.1,\quad \alpha_2^{K^*(1680)}=0.11\pm0.1,\quad \alpha_{2,\perp}^{K^*(1680)}=0.10\pm0.1,\\
B_{1,\sigma(600)}^u&=-0.42\pm0.074,\quad B_{3,\sigma(600)}^u=-0.58\pm0.23,\\
 B_{1,\sigma(600)}^s&=-0.35\pm0.061,\quad B_{3,\sigma(600)}^s=-0.43\pm0.18.\\
 B_{1,K_0^*(1430)}&=-0.57\pm0.13,\quad B_{3,K_0^*(1430)}=-0.42\pm0.22.\\
\end{split}
\end{equation}

A general fit of the parameters $\rho$ and $\phi$ to the $B\rightarrow VP$ and $B\rightarrow PV$ data indicates $X^{PV}\neq X^{VP}$, i.e. $\rho^{PV}=0.87$, $\rho^{VP}=1.07$, $\phi^{VP}=-30^0$ and $\phi^{PV}=-70^0$ \cite{Cheng:2009cn}. For the $B\rightarrow PT$ and $B\rightarrow TP$ cases, we will use the values in Ref. \cite{Cheng:2010yd}: $\rho^{TP}=0.83$, $\rho^{PT}=0.75$, $\phi^{TP}=-70^0$ and $\phi^{PT}=-30^0$. We shall assign an error of $\pm0.1$ to $\rho^{M_1M_2}$ and $\pm20^0$ to $\phi^{M_1M_2}$ for estimation of theoretical uncertainties. However, for the $B\rightarrow PS$ and $B\rightarrow SP$ decays, there is little experimental data so the values of $\rho_S$ and $\phi_S$ are not determined well, to make an estimation about $\mathcal{A_{CP}}(B^-\rightarrow K^- \sigma)$ and $\mathcal{B}(B^-\rightarrow K^-\sigma)$, we will adopt $X^{PS}=X^{SP}=(1+\rho_S e^{i\phi_S})\ln\frac{m_B}{\Lambda_h}$ as described in Sect. ${\mathrm{\uppercase\expandafter{\romannumeral3}}}$. Now we are left with only two free parameters with all the above considerations, which are the divergence parameters $\rho_S$ and $\phi_S$ for $\mathcal{A_{CP}}(B^-\rightarrow R+NR\rightarrow K^-\pi^+\pi^-)$. By fitting the theoretical result to the experimental data $\mathcal{A_{CP}}(B^-\rightarrow K^-\pi^+\pi^-)=0.678\pm0.078\pm0.0323\pm0.007$ in the region $m_{K^-\pi^+}^2<15$ $\mathrm{GeV}^2$ and $0.08<m_{\pi^+\pi^-}^2<0.66$ $\mathrm{GeV}^2$, and varying $\phi_S$ and $\rho_S$ by 0.01 each time in the range $\phi_S\in[0,2\pi]$ and $\rho_S\in[0,8]$ \cite{Bobeth:2014rra,Ciuchini:2002uv}, i.e.  $\Delta\rho_S=0.01$ and $\Delta\phi_S=0.01$, it is found that there exist ranges of the parameters $\rho_S$ and $\phi_S$ which satisfy the above experimental data. The allowed ranges are $\phi_S \in [4.75, 5.95]$ and $\rho_S\in[4.2, 8]$. Therefore, the interference of resonances ($[\pi\pi]$ resonances including $\sigma(600)$, $\rho^0(770)$, $\omega(782)$ mesons, $[K\pi]$ resonances including $K^*(892)$, $K^*(1410)$, $K_0^*(1430)$, $K^*(1680)$ and $K_2^*(1430)$ mesons) together with the nonresonance contribution can indeed induce the data for the localized $CP$ asymmetry in the $B^-\rightarrow K^-\pi^+\pi^-$ decay. It is noted that the range of $\rho_S\in[4.2, 8]$ is larger than the previously conservative choice of $\rho\leq1$ \cite{Beneke:2001ev,Beneke:2003zv}. Since the QCDF itself cannot give information about the parameters $\rho$ and $\phi$, there is no reason to restrict $\rho$ to the range $\rho\leq1$ \cite{Cheng:2009cn,Chang:2014rla,Sun:2014tfa,Zhu:2011mm}. In the pQCD approach, the possible un-negligible large weak annihilation contributions were noticed first in Refs. \cite{Keum:2000wi,Lu:2000em}. In fact, there are many experimental studies which have been successfully carried out at $B$ factories (BABAR and Belle), Tevatron (CDF and D0) and LHCb in the past and will be continued at LHCb and Belle experiments. These experiments provide highly fertile ground for theoretical studies and have yielded many exciting and important results, such as measurements of pure annihilation $B_s\rightarrow \pi \pi$ and $B_d\rightarrow K K$ decays reported recently by CDF, LHCb and Belle \cite{Aaltonen:2011jv,Aaij:2012as,Duh:2012ie}, which suggest the existence of unexpected large annihilation contributions and have attracted much attention \cite{Xiao:2011tx,Gronau:2012gs,Chang:2014rla}. Thus larger values of $\rho_S$ are acceptable when dealing with the divergence problems for $B\rightarrow SP(PS)$ decays. With the large values of $\rho_S$, it is certain that both the weak annihilation and the hard spectator scattering processes can make large contributions to $B^-\rightarrow K^- \sigma$ decays. Much more experimental and theoretical efforts are expected to understand the underlying QCD dynamics of annihilation and spectator scattering contributions. In the obtained allowed ranges for $\rho_S$ and $\phi_S$, i.e. $\rho_S\in[4.2, 8]$ and $\phi_S \in [4.75, 5.95]$, we calculate the $CP$ asymmetry parameter and the branching fraction for the $B^-\rightarrow K^- \sigma$ decay modes using Eqs. (\ref{CP asymmetry})-(\ref{e2}). The results are plotted in Figs. \ref{p1} and \ref{p2} as functions of $\rho_S$ and $\phi_S$. From these two figures and our calculated data, we obtain the predictions that $\mathcal{A_{CP}} (B^-\rightarrow K^-\sigma)\in [-0.094, -0.034]$ and $\mathcal{B}(B^-\rightarrow K^-\sigma)\in [1.82, 20.0]\times10^{-5}$ when $\rho_S$ and $\phi_S$ vary in their allowed ranges. Moreover, with the obtained values of $\rho_S$ and $\phi_S$, we can also get the localized $CP$ asymmetry $\mathcal{A_{CP}}(B^-\rightarrow K^-\pi^+\pi^-)$ induced by only $[\pi\pi]$ and only $[K\pi]$ resonances, respectively, in the same region $m_{K^-\pi^+}^2<15$ $\mathrm{GeV}^2$ and $0.08<m_{\pi^+\pi^-}^2<0.66$ $\mathrm{GeV}^2$. Inserting Eqs. (\ref{Apipi}) and (\ref{AKpi}) into Eq. (\ref{localized CP}) respectively, the results are $\mathcal{A_{CP}}(B^-\rightarrow [K^-\pi^+] \pi^- \rightarrow K^-\pi^+\pi^-)=0.086\pm0.021$ and $\mathcal{A_{CP}}(B^-\rightarrow K^-[\pi^+\pi^-] \rightarrow K^-\pi^+\pi^-)=0.585\pm0.045$. Comparing these two results, we can see the contribution from the $[K\pi]$ resonances are much smaller than that from the $[\pi\pi]$ resonances. This is because $B^-\rightarrow [K^-\pi^+] \pi$ decays are mediated by the $b\rightarrow s$ loop (penguin) transition without the $b\rightarrow u$ tree component as shown in Eqs. (\ref{Arhoomega}, \ref{Asigma}-\ref{AK2}) and also because the resonance regions of $[K\pi]$ channel mesons have smaller widths and are further away from $[\pi\pi]$ channel mesons ($\rho$, $\omega$ and $\sigma$). Therefore, the contributions from the $[K\pi]$ channel resonances are much smaller than that from $[\pi\pi]$ channel resonances. Furthermore, using  Eqs. (\ref{ANR}-\ref{x}) and Eq. (\ref{localized CP}), we also get that the nonresonance contribution as $\mathcal{A_{CP}}^{NR}(B^-\rightarrow K^-\pi^+\pi^-)=0.061\pm0.0042$ which is also much smaller than that from the $[\pi\pi]$ resonances in our studied region $m_{K^-\pi^+}^2<15$ $\mathrm{GeV}^2$ and $0.08<m_{\pi^+\pi^-}^2<0.66$ $\mathrm{GeV}^2$. Since both $\mathcal{A_{CP}}(B^-\rightarrow [K^-\pi^+]\pi^- \rightarrow K^-\pi^+\pi^-)$ and $\mathcal{A_{CP}}^{NR}(B^-\rightarrow K^-\pi^+\pi^-)$ are much smaller than $\mathcal{A_{CP}}(B^-\rightarrow K^-[\pi^+\pi^-] \rightarrow K^-\pi^+\pi^-)$. We conclude that the large localized $CP$ asymmetry $\mathcal{A_{CP}}(B^-\rightarrow K^-\pi^+\pi^-)=0.678\pm0.078\pm0.0323\pm0.007$ is mainly induced by the contributions from the $[\pi\pi]$ channel resonances.

\section{SUMMARY}
In this work, within a quasi two-body QCD factorization approach, we study the localized integrated $CP$ violation in the $B^-\rightarrow K^-\pi^+\pi^-$ decay in the region $m_{K^-\pi^+}^2<15$ $\mathrm{GeV}^2$ and $0.08<m_{\pi^+\pi^-}^2<0.66$ $\mathrm{GeV}^2$ by including the contributions from both resonances including $\sigma(600)$, $\rho^0(770)$ and $\omega(782)$ mesons from $[\pi\pi]$ channel and $K^*(892)$, $K^*(1410)$, $K_0^*(1430)$, $K^*(1680)$ and $K_2^*(1430)$ mesons from $[K\pi]$ channel. By fitting the experimental data $\mathcal{A_{CP}}(B^-\rightarrow K^-\pi^+\pi^-)=0.678\pm0.078\pm0.0323\pm0.007$ in above experimental region, it is found that there exist ranges of parameters $\rho_S$ and $\phi_S$ which satisfy the above experimental data. Thus, the resonance and nonresonance contributions can indeed induce the data for the localized $CP$ asymmetry in the $B^-\rightarrow K^-\pi^+\pi^-$ decay. The allowed ranges for $\phi_S$ and $\rho_S$ are $\phi_S \in [4.75, 5.95]$ and $\rho_S\in[4.2, 8]$ is larger than the previously conservative choice of $\rho\leq1$. In fact, there is no reason to restrict $\rho$ to the range $\rho\leq1$ because the QCDF itself cannot give information and constraint on the parameter $\rho$ and it can only be obtained through the experimental data. Large values of $\rho_S$ reveal that the contributions from the weak annihilation and the hard spectator scattering processes are both large for the $B^-\rightarrow K^- \pi^+\pi^-$ decay.  Especially, the contribution from the weak annihilation part should not be neglected.  In fact, the large weak annihilation contributions have been observed and predicted in experimental and theoretical studies. So the larger values of $\rho_S$ are acceptable when dealing with the divergence problems for the $B\rightarrow SP(PS)$ decays. With the obtained allowed ranges for $\rho_S$ and $\phi_S$, we predict the $CP$ asymmetry parameter and the branching fraction for $B^-\rightarrow K^- \sigma$. The results are $\mathcal{A_{CP}} (B^-\rightarrow K^-\sigma)\in [-0.094, -0.034]$ and $\mathcal{B}(B^-\rightarrow K^-\sigma)\in [1.82, 20.0]\times10^{-5}$ when $\rho_S$ and $\phi_S$ vary in their allowed ranges, respectively. In addition, we also calculate the localized $CP$ asymmetry $\mathcal{A_{CP}}(B^-\rightarrow K^-\pi^+\pi^-)$  only considering the $[\pi\pi]$, $[K\pi]$ resonances and nonresonance, respectively, in the same region $m_{K^-\pi^+}^2<15$ $\mathrm{GeV}^2$ and $0.08<m_{\pi^+\pi^-}^2<0.66$ $\mathrm{GeV}^2$. The results are $\mathcal{A_{CP}}(B^-\rightarrow [K^-\pi^+] \pi \rightarrow K^-\pi^+\pi^-)=0.086\pm0.021$, $\mathcal{A_{CP}}(B^-\rightarrow K^-[\pi^+\pi^-] \rightarrow K^-\pi^+\pi^-)=0.585\pm0.045$ and $\mathcal{A_{CP}}^{NR}(B^-\rightarrow K^-\pi^+\pi^-)=0.061\pm0.0042$, respectively. Therefore, the large localized $CP$ asymmetry in the $B^-\rightarrow K^-\pi^+\pi^-$ is mainly induced by the contributions from the $[\pi\pi]$ channel resonances and the contributions from $[K\pi]$ channel resonances and nonresonance are very small.

\acknowledgments
One of the authors (J.-J. Qi) is very grateful to Professor Hai-Yang Cheng, Professor Hsiang-nan Li and Professor Zhi-Gang Wang for valuable discussions. This work was supported by National Natural Science Foundation of China (Projects No. 11575023, No. 11775024, No. 11705081 and No. 11881240256).

\end{document}